\documentclass[aps,pra,twocolumn,reprint,floatfix, superscriptaddress]{revtex4-1}

\usepackage[normalem]{ulem}
\usepackage[utf8]{inputenc}
\usepackage{graphicx,amsmath,amssymb,braket}
\usepackage{color}

\definecolor{darkgreen}{rgb}{0.0, 0.4, 0.26}

\usepackage[colorlinks=true,citecolor=myblue,linkcolor=myred]{hyperref}

\definecolor{mygrey}{gray}{0.35}
\definecolor{myblue}{rgb}{0.2,0.2,0.8}
\definecolor{myzard}{cmyk}{0,0,0.05,0}
\definecolor{mywhite}{rgb}{1,1,1}
\definecolor{mywhite}{rgb}{1,1,1}
\definecolor{myred}{rgb}{1,0.,0.3}

\def\be{\begin{equation}}
\def\ee{\end{equation}}
\def\ba{\begin{align}}
\def\enda{\end{align}}
\def\bi{\begin{itemize}}
\def\ei{\end{itemize}}

\def\beq{\begin{equation}}
\def\beq{\begin{equation}}
\def\eeq{\end{equation}}

\def \rrangle{\rangle\rangle}

\newcommand\Tr{\mathrm{Tr}}
\newcommand{\mean}[1]{\langle #1\rangle}

\newcommand{\nbth}{n_b^{\mathrm{th}}}

\begin{document}

\title{Frequency-resolved photon correlations in cavity optomechanics}

\author{M.~K.~Schmidt}
\affiliation{Macquarie University Reserach Centre for Quantum Engineering (MQCEQ), MQ Photonics Research Centre, Department of Physics and Astronomy, Macquarie University, NSW 2109, Australia}
\affiliation{Materials Physics Center, CSIC-UPV/EHU, 20018 Donostia-San Sebasti\'{a}n, Spain}
\affiliation{Donostia International Physics Center, 20018 Donostia-San Sebasti\'{a}n, Spain}
\email{mikolaj.schmidt@mq.edu.au}

\author{R.~Esteban}
\affiliation{Materials Physics Center, CSIC-UPV/EHU, 20018 Donostia-San Sebasti\'{a}n, Spain}
\affiliation{Donostia International Physics Center, 20018 Donostia-San Sebasti\'{a}n, Spain}

\author{G.~Giedke}
\affiliation{Donostia International Physics Center, 20018 Donostia-San Sebasti\'{a}n, Spain}
\affiliation{IKERBASQUE, Basque Foundation for Science, 48013 Bilbao, Spain}

\author{J.~Aizpurua}
\affiliation{Materials Physics Center, CSIC-UPV/EHU, 20018 Donostia-San Sebasti\'{a}n, Spain}
\affiliation{Donostia International Physics Center, 20018 Donostia-San Sebasti\'{a}n, Spain}

\author{A.~Gonz\'alez-Tudela}
\affiliation{Institute of Fundamental Physics IFF-CSIC, Calle Serrano 113b, 28006 Madrid, Spain.}
\email{a.gonzalez.tudela@csic.es}

\begin{abstract}
	Frequency-resolved photon correlations have proven to be a useful resource to unveil nonlinearities hidden in standard observables such as the spectrum or the standard (color-blind) photon correlations. In this manuscript, we analyze the frequency-resolved correlations of the photons being emitted from an \emph{optomechanical} system where light is nonlinearly coupled to the quantized motion of a mechanical mode of a resonator, but where the quantum nonlinear response is typically hard to evidence. We present and unravel a rich landscape of frequency-resolved correlations, and discuss how the time-delayed correlations can reveal information about the dynamics of the system. 
	We also study the dependence of correlations on relevant parameters such as the single-photon coupling strength, the filtering linewidth, or the thermal noise in the environment. This enriched understanding of the system can trigger new experiments to probe nonlinear phenomena in optomechanics, and provide insights into dynamics of generic nonlinear systems. 
\end{abstract}

\maketitle

\section{Introduction}

As light emerges from an open system, it carries a lot of information about the system and its dynamics. It is up to our ingenuity to learn how to extract that information. For example, by counting the number of photons at a given frequency $\omega$~\cite{eberly77a} using a photodetector with spectral resolution $\Gamma$, we can obtain the emission spectrum $S_\Gamma(\omega)$, and extract information about the underlying level structure of the system. If instead we perform a Hanbury Brown-Twiss (HBT) experiment~\cite{hanburybrown56a}, splitting the emitted light into two beams and measuring their intensity correlation, we can measure its second-order coherence $g^{(2)}(\tau)$~\cite{glauber63a,glauber63b} which informs us about the statistical nature of the emitted light, e.g., whether it is of a quantum or classical character. These two observables ($S_\Gamma(\omega)$ and $g^{(2)}(\tau)$) are arguably the most fundamental ones to characterize open quantum optical setups. However, sometimes the information they carry is not sufficient to unravel the dynamics of complex quantum systems --- most notably, when several processes lead to multiple emission lines with competing statistics.

\begin{figure}[hbtp]
	\centering
	\includegraphics[width=\linewidth]{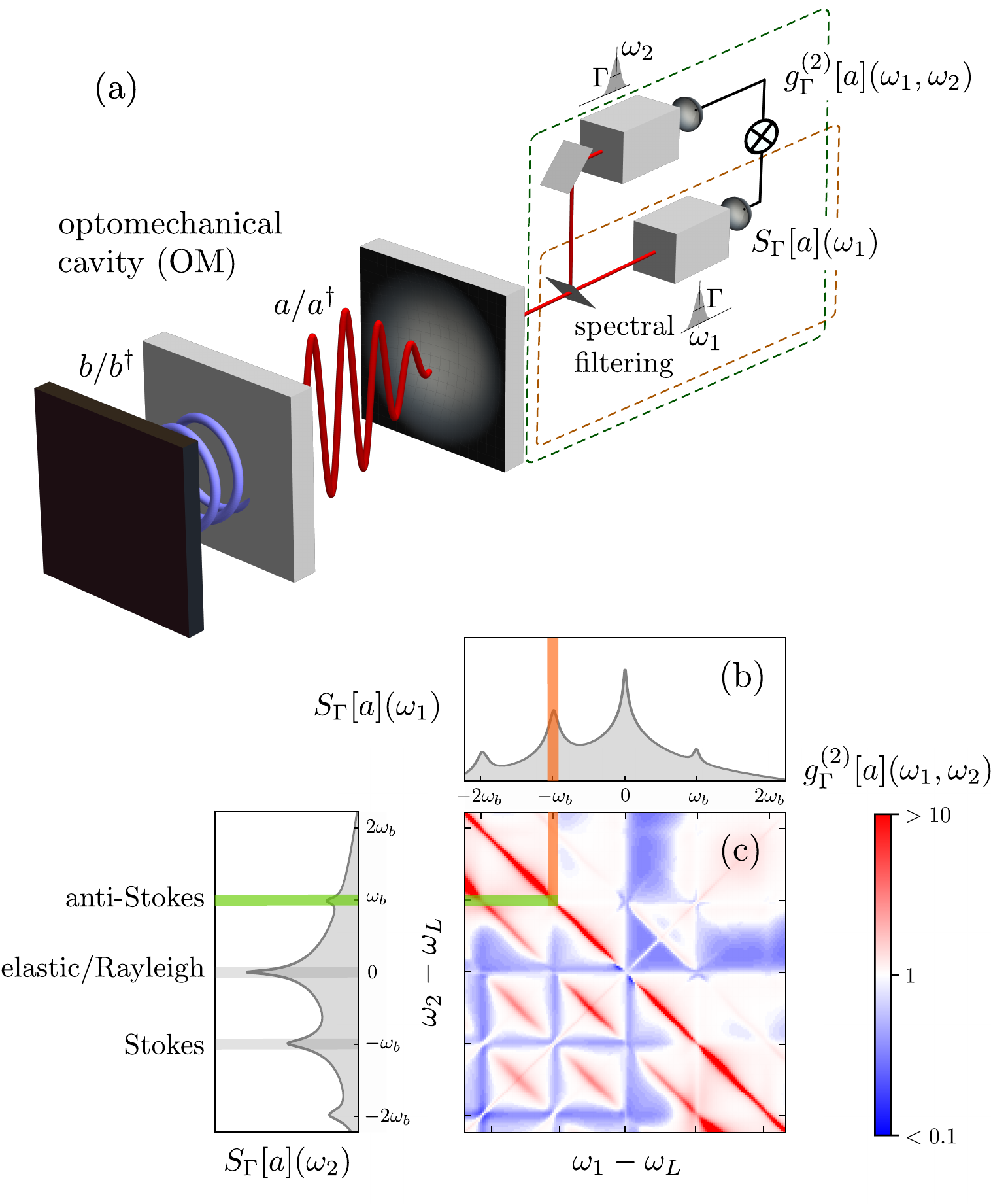}
	\caption{(a) Schematic of a generic optomechanical (OM) system in which an optical cavity (with photon creation/annihilation operators $a^\dag$/$a$) is dispersively coupled to a mechanical mode (with phonon creation/annihilation operators $b^\dag$/$b$). The spectrum of light emitted from the cavity mode $a$, $S_\Gamma[a](\omega_1)$, is measured by frequency-blind detection of the photons passing through a spectral filter tuned to frequency $\omega_1$ with resolution $\Gamma$. Directly extending this detection setup, we can measure the frequency-resolved correlations $g_\Gamma^{(2)}[a](\omega_1,\omega_2)$ by splitting the light emitted from the cavity, filtering each of the beams separately, and measuring intensity correlations of the photocurrents from the frequency-blind detectors. (b) Illustrative one-photon spectrum $S_\Gamma[a](\omega)$ of an OM system illuminated by a laser with frequency $\omega_L$, and operating in the single-photon strong coupling regime. It includes a dominant elastic scattering peak at $\omega=\omega_L$, broadened by the filter resolution $\Gamma$, and lower- and higher-frequency peaks corresponding to the phonon creation (Stokes) and annihilation (anti-Stokes) lines shifted from $\omega_L$ by multiples of the phonon frequency $\omega_b$. 
	(c) Schematic two photon spectrum (TPS; Eq.~\eqref{eq:g2freq}) of the cavity mode $a$ of an OM system, with red and blue regions denoting frequency bunched $g_\Gamma^{(2)}[a](\omega_1,\omega_2)>1$ and antibunched $g_\Gamma^{(2)}[a](\omega_1,\omega_2)<1$ emission regions, respectively. Throughout this work, the spectra and color maps in TPSs are given in logarithmic scale.}
	\label{fig:1}
\end{figure}

One of the additional tools at our disposal is the frequency-resolved version of the standard two-photon correlation function, $g_{\Gamma}^{(2)}(\omega_1,\omega_2;\tau)$, implemented by adding two frequency filters of linewidth $\Gamma$, at frequencies $\omega_{1,2}$, in each of the paths of the HBT setup \cite{nienhuis83a,cresser87a,knoll86a,nienhuis93a,nienhuis93b} (see the schematic in Fig.~\ref{fig:1}(a)). 
Originally, frequency-resolved correlations were only studied in resonance fluorescence and for particular frequency pairs, as their computation was found to be exceedingly cumbersome for more complex systems ~\cite{nienhuis83a,cresser87a,knoll86a,nienhuis93a,nienhuis93b}. However, recent theoretical developments~\cite{delvalle12a,shatokhin16a,lopezcarreno18a,holdawaya18a} triggered by the work of del Valle and co-authors~\cite{delvalle12a} have simplified this framework, and enabled the computation of full frequency correlation maps $g_{\Gamma}^{(2)}(\omega_1,\omega_2;\tau=0)$, labelled as \textit{two-photon spectra} (TPS), in a number of more complex  systems~\cite{gonzaleztudela13a,delvalle13a,sanchezmunoz15a}. Remarkably, the TPS can unveil nonlinear processes hidden in standard observables~\cite{gonzaleztudela13a,delvalle13a,sanchezmunoz15a}, and have been instrumental in inspiring novel sources of quantum light~\cite{sanchezmunoz14a,sanchezmunoz18a,sanchezmunoz14b,lopezcarreno17a} or spectroscopy techniques~\cite{lopezcarreno15a,sanchezmunoz20a}. These tantalizing prospects have boosted experimental progress on the topic, resulting already in the observation of the TPS of several systems~\cite{peiris15a,silva16a,peiris17a}.

An intriguing system that is known to exhibit very rich physics, but whose TPS has not yet been considered, is single-mode cavity optomechanics (OM)~\cite{aspelmeyer14a}, in which the optical and motional degrees of freedom of a resonator are nonlinearly coupled (see schematic in Fig.~\ref{fig:1}(a)). Cavity OM systems are particularly interesting as a platform for studying frequency-resolved intensity correlations, since the typical OM emission spectrum includes several lines from competing processes involving the creation or annihilation of vibrational quanta --- phonons (Fig.~\ref{fig:1}(b)). Intensity correlations between such processes have been used for the heralded generation of single phonons~\cite{galland14a,hong17a,velez19a,anderson18a,marinkovic18a,velez19b}. However, the theoretical descriptions of these correlations are based on simplified models~\cite{PhysRevLett.119.193603}, and provide a limited picture of the complex landscape of frequency-resolved correlations. Besides, these experiments have been performed in the \emph{linear} regime, where the nonlinearity of the coupling is removed by strongly driving the cavity. Thus, finding signatures of the nonlinear OM couplings~\cite{rabl11a,nunnenkamp11a,kronwald13a} is still an open challenge of the field, that could open the path to many OM quantum applications.


In this manuscript, we present the first complete analysis of the emission statistics from a generic OM system by studying its frequency-resolved photon correlations $g_{\Gamma}^{(2)}(\omega_1,\omega_2;\tau)$, and identifying its features with the system's underlying processes: (i) the effective Kerr cavity nonlinearity induced by the photon-phonon interaction, and (ii) a family of higher-order terms defining multi-phonon transitions, linear in the optical degree of freedom.
We study the evolution of these features in terms of both the parameters of the OM system, e.g., the optomechanical single-photon coupling $g_0/\kappa$, as well as the characteristics of the external measurement setup, e.g., the frequency filter linewidth $\Gamma$. We also calculate the temporal dynamics of frequency-resolved correlations $g_{\Gamma}^{(2)}(\omega_1,\omega_2;\tau)$, demonstrate how they encode information about the nature of emission processes, and discuss the relationship between the spectral and temporal resolution of the measurement setup. Finally, we also show how some of these frequency regions can be associated with the emission of non-classical light by studying the violation of the Cauchy-Schwarz inequality~\cite{clauser69a,sanchezmunoz14a}.

The manuscript is structured as follows. In Section~\ref{sec:theory}, we introduce the theoretical foundations of the paper, introducing the single-mode cavity OM Hamiltonian in Section~\ref{subsec:OM}, and defining the spectra $S_\Gamma(\omega)$ and frequency-resolved correlations $g_{\Gamma}^{(2)}(\omega_1,\omega_2;\tau)$ in Section \ref{subsec:observables}. In Section~\ref{sec:nonlinear} we study the TPS of two elementary nonlinear Hamiltonians, namely, a coherently driven Kerr cavity and cavity-multi-phonon interactions, which are instrumental for understanding the frequency-resolved correlations of the cavity OM Hamiltonian characterized in Section~\ref{sec:g2OM}. Finally, in Section~\ref{subsec:CSI} we demonstrate the violation of Cauchy-Schwarz inequalities in some frequency regions of the TPS, and summarize our findings in Section~\ref{sec:conclu}.
 
\section{Theoretical framework: setup and observables\label{sec:theory}}

\subsection{Single-mode cavity optomechanics \label{subsec:OM}}

Single-mode cavity optomechanics studies the interaction of a single quantized mode of an optical cavity, with frequency $\omega_a$, with a quantized vibrational, or phonon, mode of frequency $\omega_b$, as schematically depicted in Fig.~\ref{fig:1}(a). The photon-phonon coupling can be implemented in various physical systems using radiation pressure~\cite{aspelmeyer14a}, the photoelastic effect~\cite{eggleton19a}, or Raman scattering in molecular systems~\cite{schmidt16a,roelli2014molecular,schmidt2017faraday}. Irrespective of the physical mechanism inducing such interaction, the OM Hamiltonian can be written as (using $\hbar=1$ throughout the manuscript):
\begin{equation}
\label{eq:hamOM}
H_\mathrm{OM}=\omega_a a^\dagger a+\omega_b b^\dagger b-g_0 a^\dagger a (b+b^\dagger)\,,
\end{equation}
where $a^\dagger (a)/b^\dagger (b)$ are the bosonic creation (annihilation) operators of the photon/phonon mode, and $g_0$ denotes the \textit{single-photon coupling} parameter. The system is typically coherently driven with a laser exciting the cavity field, described by the following Hamiltonian:
\begin{equation}
\label{eq:hamlas}
H_L(t)=i\Omega(a e^{i\omega_L t}-a^\dagger e^{-i\omega_L t})\,,
\end{equation}
where $\Omega$ is the driving amplitude and $\omega_L$ the laser frequency. In a frame rotating with $\omega_L$ the total Hamiltonian, i.e., $H(t)=H_\mathrm{OM}+H_L(t)$, becomes time-independent:
\begin{equation}
\label{eq:hamtot}
H=\Delta_a a^\dagger a+\omega_b b^\dagger b- g_0 a^\dagger a (b+b^\dagger) + i\Omega(a-a^\dag),
\end{equation}
where $\Delta_{a}=\omega_{a}-\omega_L$ is the detuning between the cavity and laser frequencies.

Importantly, neither the optical cavity nor the mechanical mode are isolated from the photonic and phononic environments, inducing dissipation into them at rates $\kappa$ and $\gamma$, respectively. To formally account for such losses, we describe the state of the OM system using a density matrix $\rho$. Assuming that the environmental timescales are much faster than the system ones (Born-Markov approximation), the dynamics of the system is then described by the following master equation~\cite{petruccione2002theory}:
\begin{equation}\label{eq:masterequation}
\frac{d\rho}{dt}=-i[H,\rho]+\frac{\kappa}{2}\mathcal{L}_a[\rho]+\frac{\gamma \left(\nbth+1\right)}{2}\mathcal{L}_b[\rho]+\frac{\gamma \nbth}{2}\mathcal{L}_{b^\dagger}[\rho],
\end{equation}
where $\mathcal{L}_O[\rho]=\left(2 O\rho O^\dagger-O^\dagger O\rho-\rho O^\dagger O\right)$ are the Gorini-Kossakowski-Sudarshan-Lindblad (GKSL) terms, and $\nbth$ is the thermal population of the phonon bath, which both increases the decay rate of phonons, and governs the rate of incoherent pumping of mechanical vibrations by the thermal environment. Importantly, this thermal phonon bath population depends on the ratio between the phonon energy $\hbar \omega_b$ and the thermal energy $k_B T$, such that it will be very small for high-frequency {optical} phonon modes, e.g., in molecular systems~\cite{schmidt16a,roelli2014molecular,Benz726,lombardi2018pulsed}. For the sake of illustration we will assume $\nbth(T)\approx 0$ throughout most of this manuscript to keep the discussion of the physics simpler, and only briefly consider the effect of thermal population on frequency-resolved correlations in Section~\ref{sec:g2OM}. We will also ensure that the optical cavity is always weakly populated $\mean{a^\dag a}\ll 1$.

Throughout this manuscript, we will always consider the \emph{sideband-resolved} regime, where $\omega_b>\kappa$, now commonplace in many cavity OM systems~\cite{aspelmeyer14a}. The naming of this limit refers to the ability of the cavity to resolve the main relevant scattering processes identified in the system (see Fig.~\ref{fig:1}(b)): \textit{elastic/Rayleigh scattering} corresponding to emission processes that involve no exchange of excitations with the mechanical mode, so that the frequency of scattered photons matches that of the incident laser $\omega_R=\omega_L$; \textit{Stokes} and \textit{anti-Stokes processes} in which the cavity photon either loses or absorbs the energy of one phonon (or an integer number $n$ of phonons) so that light is emitted at $\omega_{S/aS}=\omega_L\mp n \omega_b$. Furthermore, we will consider systems operating from the more accessible weak coupling limit  ($g_0\ll\kappa$) to the more demanding strong single-photon coupling regime ($g_0\lesssim \kappa$) approached in physical setups based on cold atoms~\cite{brennecke08a}, molecular optomechanics~\cite{schmidt16a,roelli2014molecular,Benz726,lombardi2018pulsed}, microwave micromechanics \cite{PhysRevLett.125.023601}, and phoxonic cystals~\cite{PhysRevLett.123.223602}\footnote{Fishbonelike crystals developed by Guo \textit{et al.} in Ref.~\cite{PhysRevLett.123.223602} exhibit record single-photon cooperativity $C_0\sim 200$, but operate deeply in the sideband-unresolved regime.}. 

\subsubsection*{Polaron transformation of the OM Hamiltonian: Kerr nonlinearity and cavity-multi-phonon interaction}

To gain insight into the underlying dynamics of the OM system, and the statistics of the emitted light, it is instructive to perform the polaron transformation ${U} = \exp\left[(g_0/\omega_b) a^\dag a (b^\dag-b)\right]$ \cite{nunnenkamp11a,rabl11a} on the OM Hamiltonian. This transformation decouples the photon and phonon modes in $H_{\text{OM}}$, at the expense of transforming the harmonic cavity Hamiltonian into an anharmonic one:
\begin{equation}
\label{eq:OM.diag2}
\tilde{H}_{\text{OM}} = U^\dagger H_{\text{OM}} U = \Delta_a a^\dagger a+\omega_b b^\dagger b- \Delta_g (a^\dagger a)^2\,,
\end{equation}
with Kerr nonlinearity parameter $\Delta_g = g_0^2/\omega_b$. 
The eigenstates of the transformed Hamiltonian are then defined by the photon/phonon number states $\ket{n_a,n_b}$ and have the following energies:
\begin{align}
E_{n_a,n_b}=n_b\omega_b+n_a\omega_a -\Delta_g n_a^2\,.
\end{align}
In the regime where the nonlinearity $\Delta_g$ becomes comparable to the cavity losses $\kappa$, the system cannot readily absorb two photons with the same frequency, resulting in the well-known OM photon blockade effect~\cite{nunnenkamp11a,rabl11a}. 

In the transformed picture with non-vanishing pumping $\Omega\neq 0$, the explicit interaction between photons and phonons appears in the transformed coherent pumping term which, in the frame rotating with $\omega_L$, takes on the form:
\begin{equation}
    \label{eq:coh.pumping.transformed}
    \tilde{H}_{L} = U^\dagger H_{L} U=i\Omega\left[a e^{(g_0/\omega_b) (b^\dag-b)}-a^\dag e^{-(g_0/\omega_b) (b^\dag-b)}\right],
\end{equation}
as well as in the transformed GKSL terms, which we will for now omit.
The transformed Hamiltonian $\tilde{H}_{L}$ can be broken into several contributions by expanding the exponential:
\begin{align}
    \label{eq:coh.pumping.transformed.2}
    \tilde{H}_{L} &= i\Omega\left(a -a^\dag\right)+i\Omega \frac{g_0}{\omega_b}(a+a^\dagger)(b^\dagger-b)\nonumber \\
&~~~ +i\Omega\frac{g_0^2}{2\omega_b^2}(a-a^\dagger)(b^\dagger-b)^2+\dots\nonumber \\ 
&= \sum_{n=0}^\infty i \Omega f_n \left[a -(-1)^n a^\dag\right] \left(b^\dag -b\right)^n \nonumber \\ 
&=\sum_{n=0}^\infty \tilde{H}_L^{(n)}\,,
\end{align}
where $f_n=\left({g_0}/{\omega_b}\right)^n /n!$. When $g_0/\omega_b\ll 1$, the first term of the expansion --- the standard coherent driving term --- dominates. 

We therefore see that the transformed driven OM Hamiltonian, 
$\tilde{H}=U^\dag HU$ includes two terms which can induce nonlinear dynamics: (i) the anharmonicity of the optical mode described by Hamiltonian $\tilde{H}_{\text{OM}}$, and governed by $\Delta_g=g_0^2/\omega_b$, and (ii) a series of cavity-phonon coupling terms in $\tilde{H}_{L}$, determined by parameters $f_n \propto\left({g_0}/{\omega_b}\right)^n$, and describing $n$-phonon mediated processes. Most of the literature discussing optomechanics in the single-photon strong coupling regime is focused on exploring the photon blockade effect induced by the Kerr nonlinearity $\Delta_g>\kappa$, and identified in frequency-blind correlations in the limit of $\Omega \rightarrow 0$ \cite{nunnenkamp11a,rabl11a}. Here instead, we study the frequency-resolved photon correlations and explicitly consider both sources of nonlinearities. In fact, for pedagogical purposes, we  will analyze  in Section~\ref{sec:nonlinear} separately the TPS of these two contributions, namely, of a coherently driven Kerr-cavity Hamiltonian (in Subsection~\ref{subsec:kerrg2}), and of a higher-order multiphonon terms of Eq.~\eqref{eq:coh.pumping.transformed.2} (in Subsection~\ref{subsec:antidial}), which will help us to understand the TPS features of the complete cavity-driven OM Hamiltonian studied in Section~\ref{sec:g2OM}.

\subsection{Optical observables: spectrum and correlations \label{subsec:observables}}

As mentioned in the introduction, light emitted by the cavity carries information that can be used to characterize its underlying dynamics. The simplest measurement that one can perform is to count the number of photons emitted around a given frequency $\omega$, within a frequency window $\Gamma$ determined by the resolution of the photodetector, or the spectral width of the filter set up before the color-blind detector (see schematic in Fig.~\ref{fig:1}(a)). We label this magnitude as the one-photon spectrum, and calculate it as follows~\cite{eberly77a}:
\begin{equation}
\label{eq:spec}
S_\Gamma[a](\omega)= \lim_{t\rightarrow\infty} \frac{1}{\pi} \Re \int_0^{\infty} \text{d}\tau e^{-(i\omega+\Gamma/2) \tau}\mean{a^\dagger(t+\tau) a(t)}\,.
\end{equation}
Throughout this manuscript, we always use bracket notation, i.e., $[a]$, to denote the field operator that is being measured by the detector, e.g., here the cavity mode $a$. In the above equation, $\lim_{t\rightarrow\infty}$ indicates that the dynamics of the one-photon correlator $\mean{a^\dagger(t+\tau) a(t)}$ should be calculated in the steady state. This definition also assumes a Lorentzian filter profile with linewidth $\Gamma$, which naturally broadens the emission lines. For example, elastic scattering from the cavity will no longer appear as a Dirac delta $\delta(\omega_L)$, but rather as a Lorentzian with linewidth $\Gamma$ centered at $\omega_L$ which, as we will see, sometimes masks other features in the OM spectrum. Theoretically, this elastic contribution can be removed by observing that the optical mode $a$ can be represented as the sum of its mean value and fluctuations $a = \mean{a} + \delta a$, and calculating only the spectrum of the operator $\delta a$, which we will denote as $S_\Gamma[\delta a]$. In an experiment, this removal can be achieved by self-homodyning the emitted light with the one of the driving laser \cite{PhysRevApplied.7.044002,ElenaLPR2020}.

Another widely used quantity in the characterization of quantum optical setups is the second-order coherence function, labelled as $g^{(2)}[a](\tau)$~\cite{glauber63a,glauber63b} (denoting again in the bracket the field operator being measured), and defined as:
\begin{equation}
\label{eq:g2}
g^{(2)}[a](\tau)= \lim_{t\rightarrow\infty}\frac{
	\mean{a^\dagger(t) a^\dagger (t+\tau)a(t+\tau) a(t)}}{ \mean{a^\dagger(t) a (t)}\mean{a^\dagger (t+\tau) a (t+\tau)}}.
\end{equation}
Experimentally, $g^{(2)}[a](\tau)$ is measured with a HBT setup by dividing the light emitted from the cavity with a beam splitter, and then measuring intensity correlations between the photon detection in each of the beams~\cite{hanburybrown56a}. This quantity allows one to distinguish between the classical and quantum nature of the emission. For example, the detection of $g^{(2)}[a](0)<1$ (subpoissonian) or $g^{(2)}[a](0)<g^{(2)}[a](\tau)$ (antibunched) can both only be obtained with quantum light fields~\cite{glauber63a,glauber63b}.

Importantly, $g^{(2)}[a](\tau)$ as defined in Eq.~\eqref{eq:g2} accounts for all the different emission processes occurring in the system regardless of their frequencies. In complex quantum systems, however, where several emission processes with different frequencies and statistics simultaneously occur, this results in a loss of information which can be recovered by placing frequency filters in each of the paths of the HBT configuration (see Fig.~\ref{fig:1}(a)). This upgrade results in the measurement of the frequency-resolved two-photon correlations, which can be calculated as follows~\cite{nienhuis83a,cresser87a,knoll86a,nienhuis93a,nienhuis93b}:
\begin{align}
\label{eq:g2freq}
&g_\Gamma^{(2)}[a](\omega_1,\omega_2;\tau)=\nonumber\\
&=
\lim_{t\rightarrow\infty} \frac{\mean{:\mathcal{T}\left[A^\dagger_{\omega_1,\Gamma}(t)A^\dagger_{\omega_2,\Gamma}(t+\tau)A_{\omega_2,\Gamma}(t+\tau)A_{\omega_1,\Gamma}(t)\right] :}}{S_\Gamma[a](\omega_1)S_\Gamma[a](\omega_2)}\,.
\end{align}
Here, $A_{\omega_1,\Gamma}(t)=\int_{-\infty}^{t} \text{d}s~e^{(i\omega_1-\Gamma/2)(t-s)} a(s)$ are the operators describing light passing through the Lorentzian frequency filters, and $\mathcal{T}$ and $:\,:$ enforce the time- and normal-ordering of the cavity mode operators $a$. For $\tau=0$, this magnitude defines the two-photon spectrum (TPS)~\cite{gonzaleztudela13a,delvalle13a,sanchezmunoz15a} $g_\Gamma^{(2)}[a](\omega_1,\omega_2;\tau=0)\equiv g_\Gamma^{(2)}[a](\omega_1,\omega_2)$, which carries information about the correlations of the photons emitted at frequencies $\omega_1$ and $\omega_2$, given a filter linewidth $\Gamma$ (see Fig.~\ref{fig:1}(c) for an example of a TPS from an OM system). Generalizing the notation inherited from the standard photon correlations, we will refer to the emission with $g_\Gamma^{(2)}[a](\omega_1,\omega_2)>1$ and $g_\Gamma^{(2)}[a](\omega_1,\omega_2)<1$ as \textit{frequency bunched} and \textit{antibunched}, respectively \footnote{However, as we discuss in more detail in Section~\ref{subsec:CSI},  frequency-resolved correlations by themselves do not carry unambiguous information about the classical or quantum statistics of the emitted light, such that frequency antibunching should not be directly identified with quantum light emission.}. As shown in other works \cite{delvalle12a,gonzaleztudela13a,delvalle13a,sanchezmunoz15a}, the TPS is typically characterized by a grid of horizontal and vertical features, crossed by antidiagonal ones. The latter corresponds to filtering frequencies corresponding to two-photon processes in which the intermediate state is \emph{virtual}, i.e., not an eigenstate of the system, dubbed in other works as \emph{leapfrog} processes~\cite{gonzaleztudela13a}. The former (vertical/horizontal structure) corresponds to fixing one of the filters at a transition frequency between the eigenfrequency of the system. In the limit of very large filter linewidths, as expected, we recover the standard colorblind correlation measurements, that is $g_\infty^{(2)}[a](\omega_1,\omega_2;\tau)=g^{(2)}[a](\tau)$. 

Similarly to what it occurs for $S_\Gamma[a](\omega)$, the frequency-resolved photon correlations near the elastic scattering frequencies might also be dominated by those of the laser light $g_\Gamma^{(2)}[a](\omega_1\approx \omega_L,\omega_2\approx \omega_L)\approx 1$. Thus, in order to unveil the intrinsic dynamics of the OM interaction Hamiltonian, we will --- when explicitly noted --- consider both the TPS of the mode $a$, and of the fluctuations $\delta a$. The latter will be denoted as $g^{(2)}_\Gamma[\delta a](\omega_1,\omega_2)$, and calculated from Eq.~\eqref{eq:g2freq} by replacing the operator $a$ with $\delta a$. 
In an experiment, this measurement could be performed though the extension of the self-homodyning setup described in Ref.~\cite{PhysRevApplied.7.044002}, where the light emitted from the cavity is mixed with the driving laser before splitting it in the HBT setup. Another option to remove the elastic spectral components is to use notch filters. However, to model the effect of these filters properly, one would need to extend the formalism used here, suited for Lorentzian filters, e.g., by adopting the approach developed by Kamide \textit{et al.}~\cite{PhysRevA.92.033833}, which lies beyond the scope of this work.

The numerical framework for calculating the TPS is based on the contributions from del Valle \textit{et al.} \cite{delvalle12a} and Holdaway \textit{et al.} \cite{holdawaya18a}, and described in more detail in Appendix~\ref{sec:methods}.

\section{Correlations of underlying nonlinear processes} \label{sec:nonlinear}

In Section~\ref{sec:theory} we have shown how, using the polaron-transformed picture, the OM Hamiltonian can be mapped to that describing a pair of decoupled harmonic ($b$) and anharmonic ($a$) oscillators (see Eq.~\eqref{eq:OM.diag2}), whereas the cavity driving Hamiltonian can be expanded into a series of terms which include the standard coherent drive, but also  higher-order nonlinear interaction terms (see Eq.~\eqref{eq:coh.pumping.transformed.2}). Since all of these processes contribute to the frequency-resolved correlations of the full OM Hamiltonian, in this section we consider the frequency correlations induced by each of these individual nonlinear processes separately. This will help us to understand the complete picture when we analyze it in Section~\ref{sec:g2OM}.

Before we continue, we should note that throughout this section we discuss the spectra and correlations of mode $a$ for the respective Hamiltonians expressed through this operator, as if these Hamiltonians were given in an untransformed picture. 
Similarly, we will consider the GSKL term in the original form, given by the second term on the right-hand side of Eq.~\eqref{eq:masterequation}.

\subsection{Coherently driven Kerr system}
\label{subsec:kerrg2}

The first ingredient of the transformed optomechanical Hamiltonian that gives rise to non-trivial correlations is the phonon-mediated Kerr cavity interaction. Its Hamiltonian reads:
\begin{align}
\label{eq:Kerr.hamtot}
H_{\text{Kerr}} &= \Delta_a a^\dagger a - \Delta_g (a^\dagger a)^2 \nonumber \\ 
&= (\Delta_a-\Delta_g) a^\dagger a - \Delta_g (a^\dagger)^2 a^2 \,,
\end{align}
which, to account for losses, has to be complemented with the aforementioned GKSL term ${\kappa}/{2}\mathcal{L}_a[\rho]$. Furthermore, we assume that the cavity is driven by a coherent laser described by the Hamiltonian $H^{(0)}=-i\Omega(a^\dag-a)$, and define the complete Hamiltonian $H_{\text{Kerr},\Omega}=H_{\text{Kerr}}+H^{(0)}$. We note that the TPS of Kerr cavities was considered before in Ref.~\cite{ridolfo13a}, but in a very different scenario --- in the regime of very large Kerr nonlinearity, where one must redefine the observables to obtain physical results, and under incoherent cavity pumping.

Before considering directly the TPS of the coherently driven Kerr Hamiltonian, $H_{\text{Kerr},\Omega}$, we will first develop some intuition of its emergent features by expanding the Kerr nonlinearity around the fluctuations $\delta a$ of mode $a$ using $a\rightarrow \delta {a}+\alpha$ and $\alpha=\mean{a}$ \footnote{We further simplify the Hamiltonian by ensuring that $\alpha$ is real, by adding the phase $\arg(\alpha)$ to the operators $\delta a$}, arriving at:
\begin{align}
    (a^\dag)^2& a^2\rightarrow\left(\delta {a}^\dag +\alpha\right)^2\left(\delta a +\alpha\right)^2 \nonumber \\\nonumber
    =&~\alpha^4 + 2\alpha^3(\delta a^\dag  + \delta a) + 4 \alpha^2 \delta a^\dag \delta a \\\nonumber
    &+ \alpha^2 \left[\left(\delta a^\dag\right)^2 + \delta a^2\right] \\\nonumber
    &+ 2\alpha \left[\delta a^\dag \delta a^2 + \left(\delta a^\dag\right)^2 \delta a \right]\\
    &+ \left(\delta a^\dag\right)^2 \delta a^2.
\label{eq:power_series}
\end{align}
The terms in the first line of this expansion represent energy shifts and an additional coherent driving of $\delta a$. The second line describes one-mode squeezing, or degenerate parametric amplification, corresponding to the simultaneous creation or annihilation of two cavity photons by the incident laser, which results in a strong correlation of field components oscillating at frequencies $\omega_1$ and $\omega_2$ with $\omega_1+\omega_2=2\omega_L$. The third line describes cubic processes, and the fourth the Kerr nonlinearity of the fluctuations $\delta {a}$.


\begin{figure*}[tb]
	\centering
	\includegraphics[width=\linewidth]{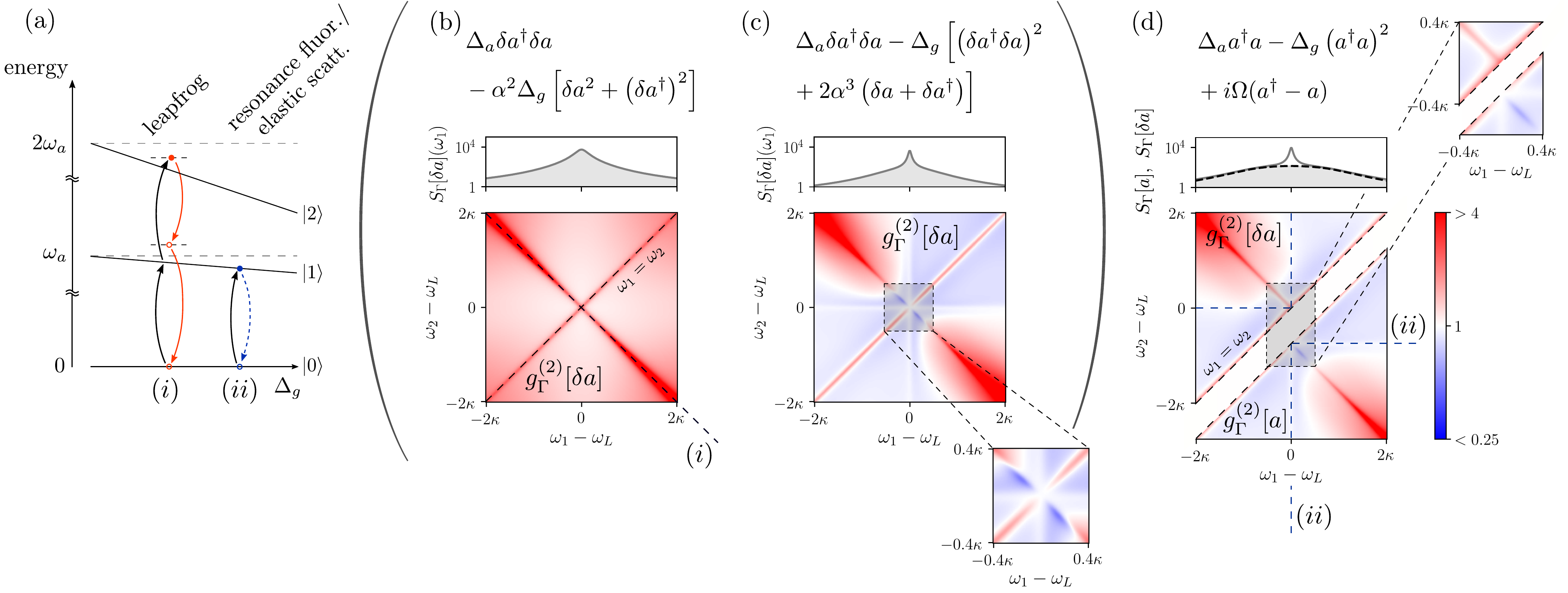}
	\caption{Spectra $S_\Gamma[a](\omega)$ (solid lines), $S_\Gamma[\delta a](\omega)$ (dashed lines), and TPS maps $g_\Gamma^{(2)}[a](\omega_1,\omega_2)$, $g_\Gamma^{(2)}[\delta a](\omega_1,\omega_2)$ of the Kerr Hamiltonian discussed in Section~\ref{subsec:kerrg2}. In (a) we show schematically the lowest energy levels of the Kerr Hamiltonian, as a function of Kerr nonlinearity, and denote \emph{(i)} two-photon \textit{leapfrog} process mediated by a virtual state and \emph{(ii)} single photon resonant fluorescence transition inducing strong emission at the laser frequency $\omega_L$. These processes are marked in the TPS in (b-d) using the same \emph{(i)} and \emph{(ii)} labels. In (b,c) we plot one- and two-photon spectra of fluctuations $\delta {a}$, for the nonlinear Hamiltonians given above the panels, including (b) one-mode squeezing of mode $\delta {a}$ and (c) effective coherent pumping and Kerr nonlinearity of $\delta {a}$. In (d) we present the spectra and correlations of both the cavity mode $a$ (solid lines in the plot of $S_\Gamma[a]$ in the upper panel, and TPS maps of $g_\Gamma^{(2)}[a]$ below the diagonal line in the lower panel), and their fluctuations $\delta a$ (dashed lines in the upper plots of $S_\Gamma[\delta a]$, and maps above the diagonal line $g_\Gamma^{(2)}[\delta a]$), in a driven Kerr system defined by Hamiltonian $H_{\text{Kerr},\Omega}=H_{\text{Kerr}}+H^{(0)}$. Insets in (c,d) show the TPS near the elastic scattering peaks. For all these systems we set $\Omega/\kappa=\Delta_g/\kappa=\Delta_a/\kappa=0.5$, filter linewidth as $\Gamma/\kappa=0.05$, decay rate of phonons as $\gamma/\kappa=0.1$, and consider laser tuned to the first excited eigenstate $\omega_L=\omega_a-\Delta_g$.}
	\label{fig:2}
\end{figure*}

In Fig.~\ref{fig:2}(b), we plot the spectra $S_\Gamma[\delta a]$ (upper panel) and TPS $g_\Gamma^{(2)}[\delta a]$ (lower panel) of the fluctuations $\delta a$ including only the terms describing the squeezing of $\delta {a}$ for a system exhibiting a strong Kerr nonlinearity $\Delta_g/\kappa=0.5$ and fixing the laser detuning to $\Delta_a=\Delta_g$~\footnote{ A different detuning choice would result in quantitatively different, but qualitatively similar features of $g^{(2)}_\Gamma[a]$ and $g^{(2)}_\Gamma[\delta a]$.}. While the one-photon spectrum shows a single Lorentzian peak, the TPS  exhibits a clear \textit{leapfrog} bunching behavior along the antidiagonal $\omega_1+\omega_2=2\omega_L$, as we predicted to occur due to the degenerate parametric amplification. We also identify \emph{indistinguishability bunching} along the diagonal $\omega_1=\omega_2$~\cite{gonzaleztudela13a,delvalle13a}, which emerges because the photons emitted within the finite-time response time of the filters ($\Gamma^{-1}$), appear as if they arrived simultaneously at the detector, resulting in an increase of the frequency-resolved bunching along this line. 

In Fig.~\ref{fig:2}(c) we plot the spectra and TPS corresponding to a nonlinear Kerr interaction between the $\delta {a}$ operators, including also the driving term $\propto \alpha^3(\delta {a}^\dag +\delta {a})$ for the same parameters as panel (b). In this case, we find that besides the leapfrog and indistinguishability bunching features, a blue region of weak frequency antibunching emerges around the central region (see the zoom in the inset panel). Its origin can be attributed to the nonlinear energy shift of the excited states due to the Kerr nonlinearity, which prevents the driving term from populating efficiently the state $\ket{2}$ (see Fig.~\ref{fig:2}(a)). The system then becomes reminiscent of a two-level system, whose TPS was discussed in Ref.~\cite{gonzaleztudela13a}, exhibiting particularly strong antibunching near the resonance $\omega_i\sim\omega_L$ along the antidiagonal $\omega_1+\omega_2=2\omega_L$ (see inset in Fig,~\ref{fig:2}(c)). In the case of frequency-blind correlations, this is equivalent to the photon blockade effect.


Finally, we plot in Fig.~\ref{fig:2}(d) the spectra and the TPS for the coherently driven Kerr Hamiltonian $H_{\text{Kerr},\Omega}$, for both the cavity mode $a$, and its fluctuations $\delta a$ (plotting each TPS separately by cutting the correlation map along the diagonal defined by $\omega_1=\omega_2$). Both the spectra and TPS of mode $a$ are nearly identical to those exhibited by the fluctuations $\delta {a}$ in Fig.~\ref{fig:2}(c). The main difference between the spectra of $a$ and $\delta a$ in Fig.~\ref{fig:2}(d) is that the latter does not include the filter-broadened elastic scattering peak. Further, the TPS of $\delta a$ for the $H_{\text{Kerr},\Omega}$ Hamiltonian does not show the blue dip on the antidiagonal
(see inset in Fig,~\ref{fig:2}(d)). This indicates that the admixing of the coherent field $\mean{a}$ to the light emerging from a Kerr system can switch the statistical properties of the leapfrog-dominated emission from strong frequency bunching to antibunching. A similar effect was recently analyzed by Casalengua \textit{et al.} in the context of frequency-blind correlations~\cite{casalengua2020tuning}.

\subsection{Multi-phonon processes}
\label{subsec:antidial}
\begin{figure*}[tb]
	\centering
	\includegraphics[width=\linewidth]{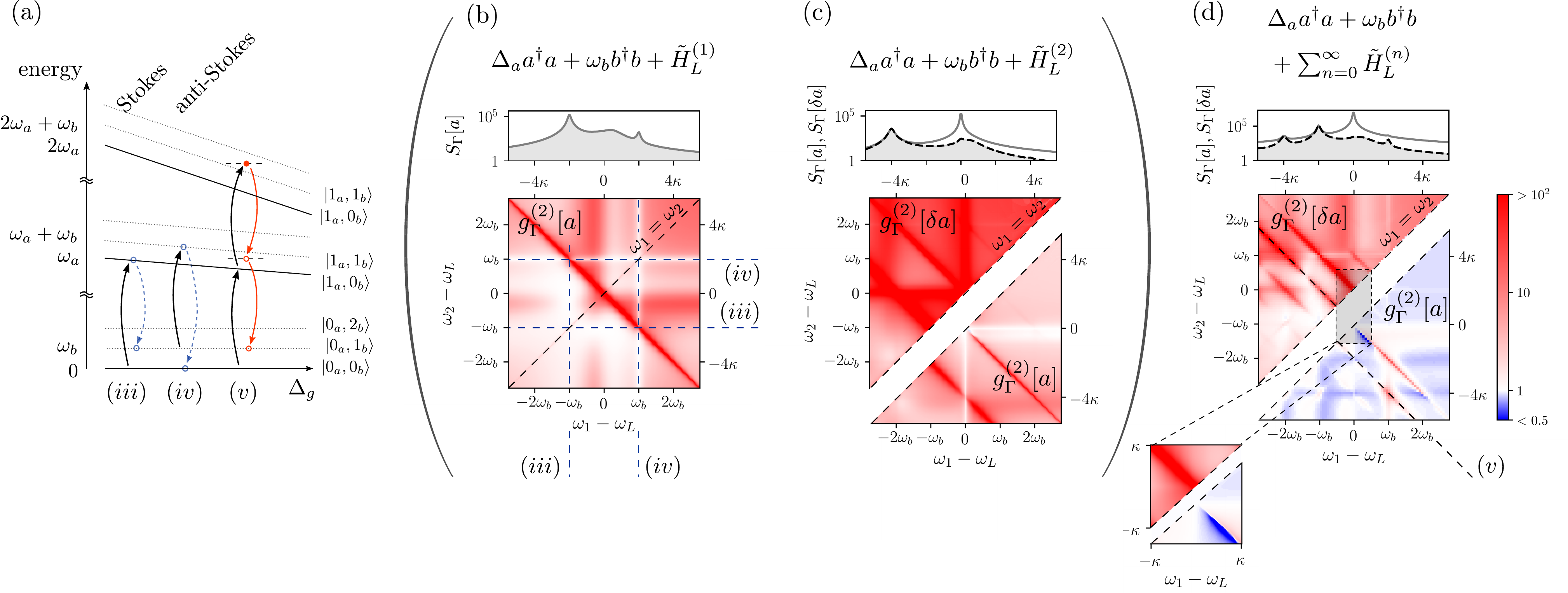}
	\caption{Spectra $S_\Gamma[a](\omega)$ (solid lines), $S_\Gamma[\delta a](\omega)$ (dashed lines), and TPS maps $g_\Gamma^{(2)}[a](\omega_1,\omega_2)$ $g_\Gamma^{(2)}[\delta a](\omega_1,\omega_2)$ of multi-phonon processes discussed in Section~\ref{subsec:antidial}. In (a) we show schematically the lowest energy levels of the multi-phonon Hamiltonian $\tilde{H}_L$, as a function of the effective Kerr nonlinearity, denoting the single photon \emph{(iii)} Stokes and \emph{(iv)} anti-Stokes emission process, as well as \emph{(v)} the two-photon leapfrog transition with the final state corresponding to an excited phonon state $\ket{0_a,1_b}$. These processes are marked in the TPS in (b-d). Panels (b,c) correspond to the results when including only particular terms of the explicitly expanded multi-phonon Hamiltonian $\tilde{H}_L$ (Eq.~\eqref{eq:coh.pumping.transformed.2}), while (d) consider the full expansion. The spectra in (c,d) exhibit a central peak corresponding to the elastically scattered laser light ($\omega_1=\omega_L$), and strongly asymmetric peaks originating from multi-phonon processes which can be assigned to particular terms in the expansion (b,c). Similarly, TPS maps include multiple antidiagonal features arising from the subsequent terms in the Hamiltonian. In (c,d), we present the spectra and correlations of both the cavity modes $a$ (solid lines in upper plots of $S_\Gamma[a]$, and maps below the diagonal line $g_\Gamma^{(2)}[a]$), and their fluctuations $\delta a$ (dashed lines in upper plots of $S_\Gamma[\delta a]$, and maps above the diagonal line $g_\Gamma^{(2)}[\delta a]$). For all these systems (b-d), we consider large single-photon coupling $g_0 = \kappa$ and coherent pumping $\Omega/\kappa=0.1$, we set the filter linewidth as $\Gamma/\kappa=0.05$, decay rate of phonons as $\gamma/\kappa=0.1$, and consider a laser tuned to the first excited eigenstate $\omega_L=\omega_a-\Delta_g$.}
	\label{fig:3}
\end{figure*}

The second nonlinear ingredient of the polaron-transformed OM Hamiltonian is described by the series of terms, written in Eq.~\eqref{eq:coh.pumping.transformed.2}, linear in the optical degree of freedom $a$, and increasingly nonlinear in the mechanical mode operator $b$. In Fig.~\ref{fig:3} we analyze how they contribute to the emergence of additional features in the spectra and TPS of the system:
\paragraph{1st term} $\tilde{H}_L^{(1)}=i\Omega g_0/\omega_b\left(a+a^\dag\right)\left(b^\dag-b\right)$. Since this Hamiltonian does not include a laser driving term, we find that $\mean{a}=0$. Thus, the one-photon spectra and TPS of the $a$ mode will be the same as the one of $\delta a$, which is what is plotted in Fig.~\ref{fig:3}(b). In that figure, we observe how the spectra $S_\Gamma[a](\omega)$ develops in this case three peaks corresponding to the elastic ($\omega\approx \omega_L$) and Stokes/anti-Stokes emission processes ($\omega\approx \omega_L\pm \omega_b$), denoted as processes \emph{(iii)} and \emph{(iv)} in Fig.~\ref{fig:3}(a).
Regarding the corresponding TPS, its most prominent feature corresponds to the strong bunching antidiagonal around $\omega_1+\omega_2\approx 2\omega_L$. This feature can be understood by separating the \textit{passive} (beamsplitter-like) $ab^\dagger+a^\dagger b$ and \textit{active} (two-mode-squeezing) $ab+a^\dagger b^\dagger$ interaction terms. In a perturbation picture, the latter will drive the system into states $\ket{i_a,i_b}$, generating strong $a \leftrightarrow b$ intensity correlations, and the former will transfer these correlations into intensity autocorrelations of mode $a$ (simultaneously, $b$), populating the state $\ket{2_a,0_b}$. This is effectively the same mechanism (quadratic in both $g_0$ and $\Omega$) as the degenerate squeezing identified in the expansion of the Kerr interaction, and thus results in the strong antidiagonal bunching line in the TPS (we discuss these features in more detail in Appendix~\ref{sec:coupled_cavities}).

On top of this strong bunching antidiagonal, the TPS
develops a vertical/horizontal grid of correlations $\approx 1$ (white regions) when fixing one of the filters to the Stokes/anti-Stokes frequencies, marked as \emph{(iii)} and \emph{(iv)} in Fig.~\ref{fig:3}(a-b). These correlations correspond to emission processes between the real energy levels of the system and will be discussed in more detail in the subsequent section.

    Finally, let us note that the Hamiltonian $\tilde{H}_L^{(1)}$ is identical to the linearized OM Hamiltonian, and describes the dynamics of OM systems operating in the single-photon weak-coupling limit ($g_0\ll \kappa$), which is where the majority of OM systems currently work. 

\paragraph{2nd term}
$\tilde{H}_L^{(2)}=i\Omega g_0^2/(2\omega_b^2)\left(a-a^\dag\right)\left(b^\dag-b\right)^2$. In Fig.~\ref{fig:3}(c) we plot the one-photon spectra and TPS of this Hamiltonian for both the cavity mode $a$ and its fluctuations $\delta a$. The one-photon spectra are both dominated by two peaks: one at the laser frequency $\omega\approx \omega_L$, and another one displaced twice the phonon frequency $\omega\approx \omega_L-2\omega_b$. This translates into a TPS developing two strong bunching antidiagonals at both $\omega_1+\omega_2\approx 2\omega_L,2\omega_L-2\omega_b$. Again, these features can be intuitively understood by expanding $\tilde{H}_L^{(2)}$ into two terms:
    \begin{equation}
        (a^\dagger - a)(b^\dagger- b)^2 = (a^\dagger - a)\left[\underbrace{(b^\dagger)^2+b^2}_{\text{\textit{sq}}}\overbrace{-2b^\dagger b-1}^\text{\textit{dis}}\right].
        \label{eq:H2_2}
    \end{equation}
    (i) In the absence of the $b$-squeezing terms (\textit{sq}) or additional driving terms, the $b$ mode population vanishes $\mean{b^\dagger b}=0$, and the dispersive term \textit{dis} turns into a coherent drive for mode $a$, decoupled from $b$, generating TPS features $\sim 1$ at $\omega_{1/2}=0$. Alternatively, in the absence of the \textit{dis} terms, this Hamiltonian becomes similar to $\tilde{H}_L^{(1)}$ with two-phonon operators replacing single-phonon operators. This explains the onset of the strong antidiagonal $\omega_1+\omega_2=2\omega_L$ and indistinguishability bunching lines along the diagonal of the TPS of Hamiltonian $\tilde{H}_L^{(2)}$ in Fig.~\ref{fig:3}(c). (ii) The two-photon Stokes emission peak in the one-photon spectra and the shifted antidiagonal $\omega_1+\omega_2=2\omega_L - 2\omega_b$ can be explained by tracking the Hilbert space accessed by the Hamiltonian $\tilde{H}_L^{(2)}$ when both the \textit{sq} and \textit{dis} terms are present:  $\ket{i_a,i_b}$ such that $i_b$ is even. This space includes both states $\ket{2_a,0_b}$ and $\ket{0_a,2_b}$, that can be connected by a two-photon processes with energy given by $\omega_1+\omega_2=2\omega_L - 2\omega_b$, which defines the off-centre diagonal observed in Fig.~\ref{fig:3}(c).\\
    Furthermore, we note that the non-vanishing coherent amplitude contribution $\alpha$ to the $a$ operator generates a strong elastic, filter-broadened emission line in the spectrum $S_\Gamma[a]$, which also homogeneously lowers the values of the observed TPS $g_\Gamma^{(2)}(\omega_1,\omega_2)$, by flooding the detection with coherent light (with values $\sim 1$). This effect can be seen by comparing the TPS correlations obtained for $a$ and for the $\delta a$ fluctuations (regions below and above the diagonal in Fig.~\ref{fig:3}(c))

\paragraph{Full Hamiltonian $\tilde{H}_L$.}
Finally, we show in Fig.~\ref{fig:3}(d) the spectra and TPS of the entire series of linear drive, single- and multi-phonon processes $\tilde{H}_L$ given in Eq.~\eqref{eq:coh.pumping.transformed.2} (see caption of Fig.~\ref{fig:3} for parameters), for both operator $a$ and its fluctuations $\delta a$. The spectrum is made up of multiple emission peaks at $\omega_i=\omega_L\pm n\omega_b$, which are also visible in the TPS as vertical and horizontal lines. We also identify a series of antidiagonal bunching features $\omega_1+\omega_2=2\omega_L+ n\omega_b$ resulting from the multi-phonon leapfrog transitions, including one ($n=-1$) depicted in panels (a) and (d) as \emph{(v)}, which describes emission into the excited vibrational state $\ket{0_a,1_b}$. Interestingly, the TPS of the cavity mode $g_\Gamma^{(2)}[a](\omega_1,\omega_2)$ and its fluctuations $g_\Gamma^{(2)}[\delta a](\omega_1,\omega_2)$ shown in Fig.~\ref{fig:3}(d) appear to be qualitatively very different. In particular, one observes that the absence of the elastic scattering contribution seemingly removes all the traces of frequency antibunched (blue) emission. This is particularly striking along the central antidiagonal line, where the strong frequency antibunching in $g_\Gamma^{(2)}[a](\omega_1,\omega_2)$ near the elastic emission peak is entirely replaced by a bunched characteristic of fluctuations $g_\Gamma^{(2)}[\delta a](\omega_1,\omega_2)$, similarly to what occurs in the TPS of the coherently driven Kerr cavity that we show in Fig.~\ref{fig:2}(d). We note that this strong contrast, and more generally, the presence of large white and blue areas in the TPS $g^{(2)}_\Gamma[a]$ in Fig.~\ref{fig:3}(d), is largely due to the linear drive included in $\tilde{H}_L$, but absent in $\tilde{H}_L^{(1,2)}$. In the presence of coherent driving, in Fig.~\ref{fig:3}(b,c) we would also find mostly white $g^{(2)}_\Gamma[a]$, with some significantly frequency antibunched (blue) areas.

\section{Frequency-resolved correlations in single-cavity optomechanics cavities}
\label{sec:g2OM}

\begin{figure*}[tb]
	\centering
	\includegraphics[width=.9\linewidth]{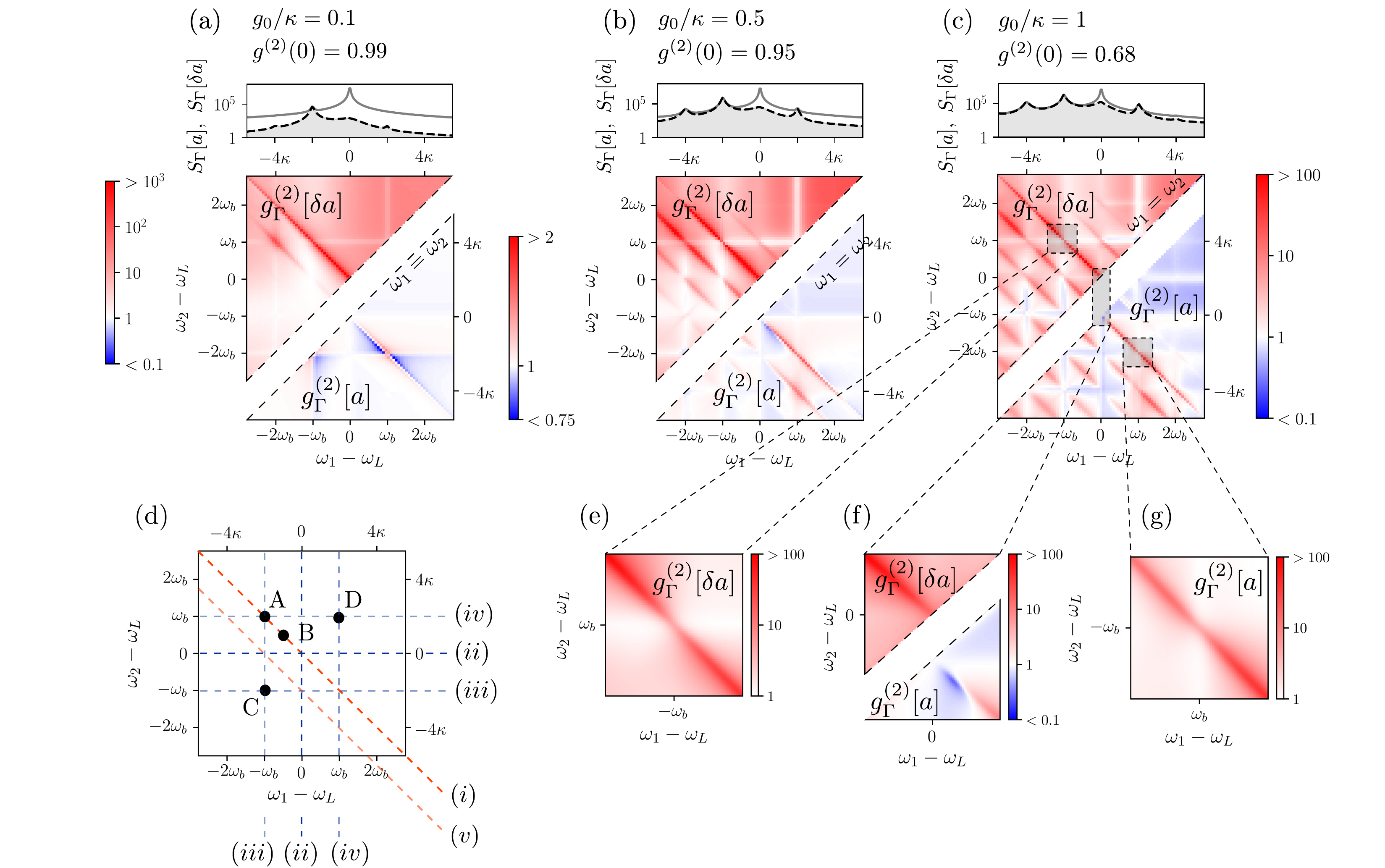}
	\caption{(a-c) Spectra $S_\Gamma[a](\omega)$ (solid), $S_\Gamma[\delta a](\omega)$ (dashed), and TPS maps $g_\Gamma^{(2)}[a](\omega_1,\omega_2)$, $g_\Gamma^{(2)}[\delta a](\omega_1,\omega_2)$ of a cavity OM system (Eqs.~(\ref{eq:hamtot},\ref{eq:masterequation})) in the sideband-resolved regime ($\omega_b/\kappa=2$), for three values of $g_0/\kappa=(0.1, 0.5, 1)$. For these parameters, the photon nonlinear shift induced by the phonons is $\Delta_g/\kappa=(0.005, 0.125, 0.5)$, and the values of the frequency-blind two-photon correlations $g^{(2)}$ given above the spectra indicate the onset of a strong photon blockade. (d) Guidelines of different processes (i-v) appearing in the TPS of OM, shown schematically in panels in Fig.~\ref{fig:2}(a) and in Fig.~\ref{fig:3}(a). We also mark as A, B, C and D the specific pairs of filter wavelengths that are discussed in the main text.	(e-g) Magnified regions of the TPS maps near the Stokes--anti-Stokes emission correlation ((e,g), $\omega_{1/2}=\omega_L\pm\omega_b$), and near the elastic scattering peaks (f), for both the fluctuations $\delta a$ and dressed cavity field $a$. For all the systems we set the filter linewidth as $\Gamma/\kappa=0.05$, decay rate of phonons as $\gamma/\kappa=0.1$, and consider a laser with amplitude $\Omega/\kappa=0.1$ tuned to the first excited eigenstate $\omega_L=\omega_a-\Delta_g$.}
	\label{fig:4}
\end{figure*}

In the previous section we studied the features of the TPS of the two terms that form the $H_\mathrm{OM}$ in the polaron picture. With this knowledge as a basis, in this section we finally consider the emission from single-cavity OM systems as described in Eqs.~\eqref{eq:hamtot} and \eqref{eq:masterequation}, and analyze their TPS in detail. We relate to the knowledge developed in the previous section when possible, and expand it when new features emerge in the correlation maps of the OM setup.

In particular, we study the dependence of the TPS on relevant parameters of the system such as optomechanical coupling and thermal phonon population, in subsection~\ref{subsec:TPS}. Then, in subsection~\ref{subsec:SaS} we consider the dynamics of the correlations with time-delay $\tau$ of some of the features of the TPS: the Stokes--anti-Stokes, and leapfrog correlations.


\subsection{One- and two-photon spectra}
\label{subsec:TPS}


In Fig.~\ref{fig:4} we plot the spectra and TPS of cavity OM setups for three different single-photon couplings strengths $g_0/\kappa=(0.1, 0.5, 1)$, corresponding to effective Kerr nonlinearities $\Delta_g/\kappa=(0.005, 0.125, 0.5)$. As in Figs.~\ref{fig:3}(c-d), we consider spectra calculated from both the cavity mode operator $a$ ($S_\Gamma[a]$ and $g_\Gamma^{(2)}[a](\omega_1,\omega_2)$), and its fluctuations around the steady-state displacement $\delta a$ ($S_\Gamma[\delta a]$ and $g_\Gamma^{(2)}[\delta a](\omega_1,\omega_2)$), displayed together again by cutting the correlation map along the diagonal. 

For the weakest coupling $g_0/\kappa=0.1$, the one-photon spectrum of mode $a$, shown in solid gray line in Fig.~\ref{fig:4}(a), is almost entirely dominated by the elastic peak, as it also occurs for the corresponding color-blind photon correlations, which display a value very close to $1$, i.e., $g^{(2)}(0)=0.99$. Interestingly, the corresponding TPS already reveals a non-trivial correlation structure with some regions of frequency bunching and antibunching. For example, the TPS features a vertical and horizontal grid of uncorrelated transitions characterized by  $g_\Gamma^{(2)}[a](\omega_1,\omega_2)\approx 1$  (marked by a grid of white lines, and including Stokes and anti-Stokes processes denoted schematically as \emph{(iii)} and \emph{(iv)} in Fig.~\ref{fig:3}(a)). Conversely, the spectrum and TPS of the fluctuations are similar to those shown and discussed in Fig.~\ref{fig:3}(b) corresponding to the Hamiltonian $\tilde{H}_L^{(1)}$: in the spectrum (dashed black line in Fig.~\ref{fig:4}(a)) the elastic scattering peak becomes suppressed, which favours the observation of the phonon sideband peaks; in the TPS, the weak frequency-antibunched regions disappear favoring the appearance of frequency-bunched regions. The most prominent features are still the horizontal/vertical grid of uncorrelated (white) Stokes/anti-Stokes transitions, and the bunched antidiagonal line at $\omega_1+\omega_2=2\omega_L$. This qualitative resemblance with the features of $\tilde{H}_L^{(1)}$ corroborates our expectation that in the regime of weak single-photon coupling $g_0/\kappa\ll 1$ analyzed in Fig.~\ref{fig:4}(a), the TPS of an OM system will be nearly identical to that described by linearizing the OM Hamiltonian. 

As we increase the coupling $g_0$ (Figs.~\ref{fig:4}(b,c)) { towards the single-photon strong coupling regime, where the linearized Hamiltonian fails to describe the response of the optomechanical system~\cite{nunnenkamp11a,rabl11a}, we begin to observe features originating from the nonlinear form of the coupling. In the spectra $S_\Gamma$, we find additional Stokes and anti-Stokes peaks, corresponding to the multi-phonon processes. In the TPS, we also identify antidiagonal bunching lines (at $\omega_1+\omega_2=2\omega_L-n\omega_b$ for $n=1,~2,~3,...$) originating from leapfrog emission driving the system \textit{to} the excited vibrational states (including processes \emph{(v)} marked in Fig.~\ref{fig:3}(a) for $n=1$), and a much weaker antidiagonal line at $\omega_1+\omega_2=2\omega_L+\omega_b$, describing emission induced by the laser driving the system \textit{from} the excited vibrational states. For the largest coupling (panel (c)), we find --- as in Fig.~\ref{fig:3} --- a weak and narrow antibunching along the central antidiagonal (see panel (f)), resulting from the interference of elastic scattering peak and leapfrog emission. We also find that some of the previously highlighted features become stronger, e.g., the antidiagonal bunching line at $\omega_1+\omega_2=2\omega_L$, as well as the frequency-antibunched (blue) regions for operator $a$.}

Finally, the TPS for both the cavity operator $a$ and the fluctuations $\delta a$ include a grid of vertical and horizontal antibunching features coming from single-photon transitions between the relevant energy levels of the system, including $\omega_i=\omega_L\pm \omega_b$ (processes \emph{(iii,~iv)} in Fig.~\ref{fig:3}(a)). Let us now highlight how the TPS look like around some relevant crossings of this grid:

\begin{itemize}

\item The Stokes--Stokes correlation (point C in Fig.~\ref{fig:4}(d)) features a structure, in both $a$ and $\delta a$, similar to one identified in Kerr cavities (see Fig.~\ref{fig:2}(d)) or two-level systems~\cite{gonzaleztudela13a}, with vertical ($\omega_1=\omega_S$) and horizontal ($\omega_2=\omega_S$) lines depicting strong antibunching, and crossing with the bunching features associated with the leapfrog $\omega_1+\omega_2=2\omega_L-2\omega_b$ and indistinguishability. This two-level-like TPS results from the effective Kerr detuning between the states connected via Stokes transitions: $\ket{0_a,0_b}$, $\ket{1_a,1_b}$ and $\ket{2_a,2_b}$. 

\item The anti-Stokes--anti-Stokes (point D) crossing is mostly dominated by the elastic response since the probability for these anti-Stokes processes to occur is very low with these system parameters. 

\item The Stokes--anti-Stokes correlations ($\omega_{1/2}=\omega_{S/aS}=\omega_L\pm\omega_b$), identified as point A in Fig.~\ref{fig:4}(d), are highlighted in Figs.~\ref{fig:4}(e,g) for the TPS of the fluctuations ($\delta a$) and cavity mode, respectively. In both cases, we clearly find that the strong bunching associated with the leapfrog process \emph{(i)} becomes significantly suppressed (note the logarithmic color scale) when the filters are tuned to the Stokes and anti-Stokes lines~\cite{PhysRevLett.119.193603}. {This change of behavior occurs because different photon-emission processes start to dominate the correlations (strongly affecting both the numerator and denominator in Eq. \ref{eq:g2}) : \textit{a) Stokes--anti-Stokes transitions mediated by a real, one-phonon state} $\ket{0_a,1_b}$, described by sequential transitions \emph{(iii)} and \emph{(iv)} depicted in Fig.~\ref{fig:3}(a). As described in Ref.~\cite{schmidt16a}, in the absence of thermal phonon population, and for sufficiently weak coherent pumping, an anti-Stokes photon is necessarily accompanied by a previous Stokes process, yielding strong bunching. \textit{b) Transitions} from a two-photon state, similar to the leapfrog process \emph{(i)}, but where the intermediate state is a proper energy level of the system,  \textit{c) Stokes and anti-Stokes transitions mediated by the thermal phonons} (when present). The contribution from these three mechanisms and, when present, interference with the elastic scattering, can give rise to a nontrivial dependence of Stokes--anti-Stokes correlations on the coupling parameter $g_0$ and the time delay $\tau$, which we discuss in detail in the following subsections.}

\end{itemize}

\subsection{Stokes--anti-Stokes and leapfrog correlations - dependence on parameters}
\label{subsec:SaS}

\begin{figure}[tb]
	\centering
	\includegraphics[width=\linewidth]{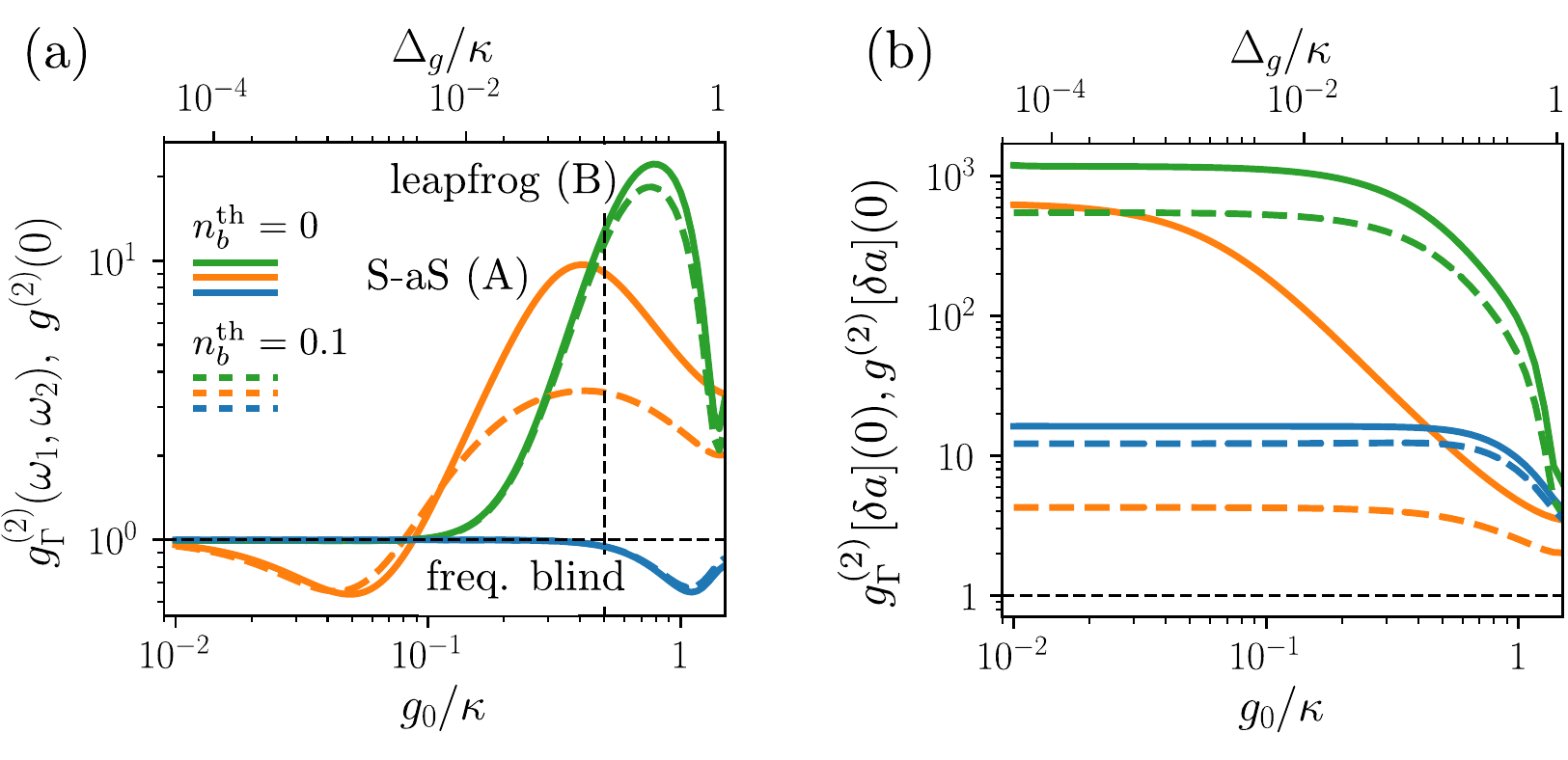}
	\caption{(a,b) Dependence of the two-photon correlations on the  single-photon coupling parameter $g_0$, calculated for (a) the cavity mode $a$ and (b) its fluctuations $\delta a$. We consider correlations of
	Stokes--anti-Stokes pairs (orange lines; $\omega_1=\omega_L-\omega_b$, $\omega_2=\omega_L+\omega_b$), leapfrog transitions (green; $\omega_1=\omega_L-0.5\omega_b$, $\omega_2=\omega_L+0.5\omega_b$) and frequency-blind correlations $g^{(2)}(\tau)$ (blue), with the solid and dashed lines denoting calculations assuming vanishing ($\nbth=0$) and non-vanishing ($\nbth=0.1$) thermal populations of phonons, respectively.  The filter width is set to $\Gamma/\kappa=0.05$.
	Besides these parameters, for all the systems we set $(\omega_b, \gamma)/\kappa=(2, 0.1)$, and consider the laser with amplitude $\Omega/\kappa=0.1$ tuned to the first excited eigenstate $\omega_L=\omega_a-\Delta_g$.
}
	\label{fig:5coupling}
\end{figure}


In this section, we will explore in more detail the dependence on the parameters of the OM system of some of the features of the TPS discussed in the previous section. In particular, we choose two points of the TPS of Fig.~\ref{fig:4}, namely, a point that corresponds to a leapfrog correlation line (B in Fig.~\ref{fig:4}(d)) and the Stokes--anti-Stokes correlations (point A in Fig.~\ref{fig:4}(d)). We plot in green/orange lines, respectively, in Fig.~\ref{fig:5coupling}(a) the evolution of their correlations as a function of the \textit{granularity parameter} $g_0/\kappa$ (the importance of the width of the filter is discussed in Appendix \ref{ap:filter}) . We also plot together with them the evolution of the colorblind photon correlation $g^{(2)}_\Gamma[a](0)$ (solid blue). The correlations $g_\Gamma^{(2)}[a](\omega_1,\omega_2)$ in this panel are obtained for the cavity operator $a$. We observe that the leapfrog correlation grows from the elastic-field induced $g_\Gamma^{(2)}[a](\omega_1,\omega_2;\tau=0)\approx 1$, to become very strongly bunched for $g_0/\kappa\sim 0.7$, and then relaxing back to 1 for $g_0/\kappa \sim 1$. On the contrary, the Stokes--anti-Stokes correlations exhibit a strong frequency antibunching for coupling parameters $g_0/\kappa \sim 0.05$, and then grows to develop a strongly bunched signal until it relaxes back to one, similarly to the leapfrog point correlations. In both cases, it is important to note the higher sensitivity to $g_0/\kappa$ of the frequency-resolved correlations compared to its colorblind counterpart, which only starts deviating from the elastically dominated correlations to show the expected OM blockade~\cite{RablPRL11,NunnenkampPRL11} for $g_0/\kappa\gtrsim 0.3$. Note that all these observations still hold if we consider a nonvanishing incoherent thermal pumping of the mechanical degree of freedom ($\nbth=0.1$, depicted by dashed lines of the same color). 

A particularly striking feature of the frequency-resolved correlations depicted in Fig.~\ref{fig:5coupling}(a) is the strong frequency antibunching displayed by the Stokes-anti Stokes correlations for small $g_0/\kappa$ parameters. To learn more about its behaviour, as we did in previous Sections, we will compare the results of Fig.~\ref{fig:5coupling}(a) with the correlations of the fluctuations $\delta a$ for the same  parameters, which is shown in Fig.~\ref{fig:5coupling}(b). There, we find no trace of frequency antibunching for any of those points, as the Stokes--anti-Stokes, leapfrog, and even the frequency-blind correlations exhibit bunching for the entire range of the coupling parameter. This calculation suggests that the frequency antibunching of the Stokes--anti-Stokes point could be related to the effective \textit{interference} between the Stokes emission, and the elastically scattered laser light, in an effect which we dub as \textit{interference antibunching}. We will confirm this intuition in the next subsection by studying in detail the time dynamics of such correlation points.



\subsection{Dynamics of frequency-resolved correlations}
\label{subsec:time}

\begin{figure}[tb]
	\centering
 	\includegraphics[width=\linewidth]{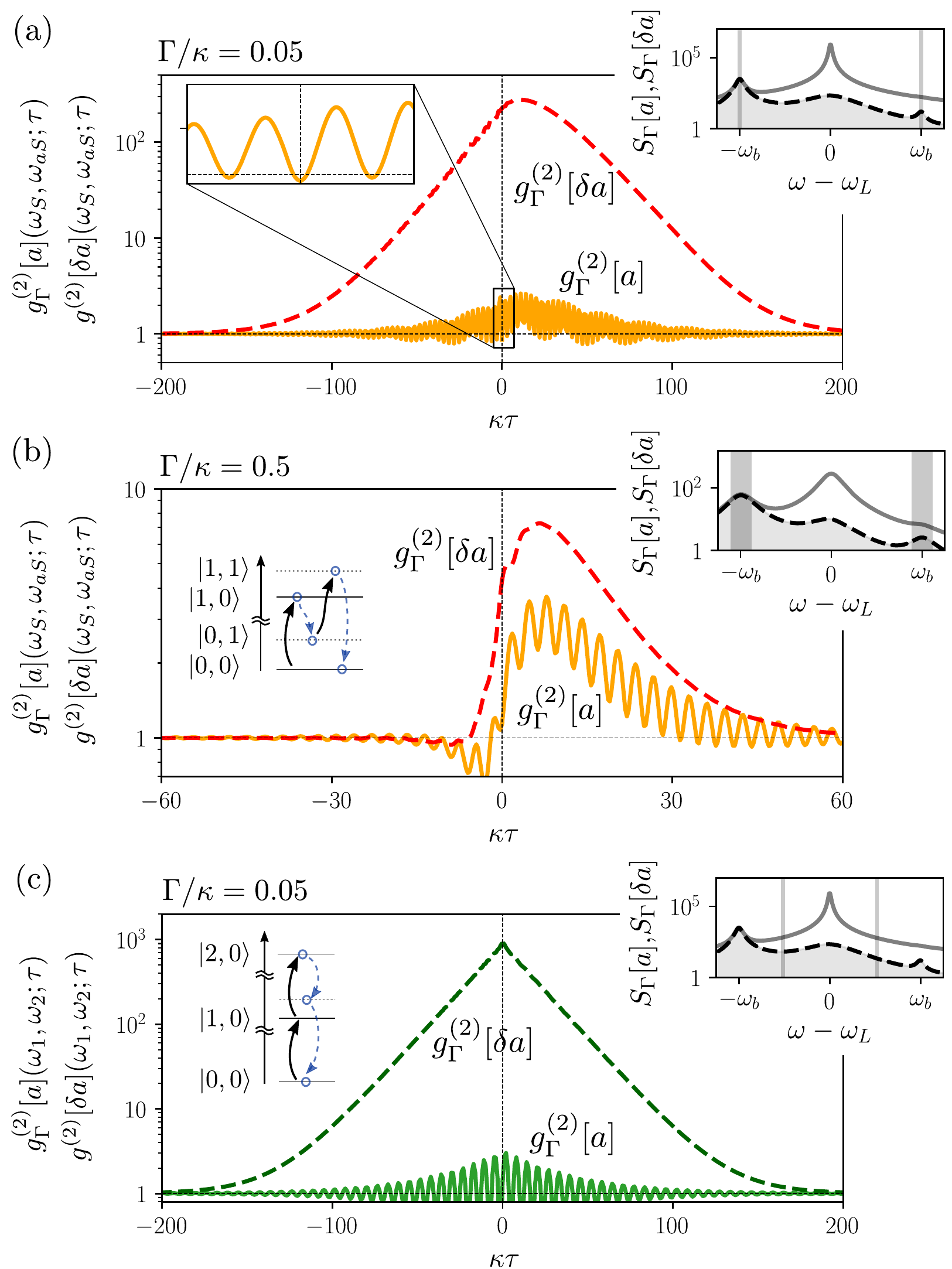}
	\caption{
	Two-photon time-delayed correlations $g_\Gamma^{(2)}[a](\omega_1,\omega_2;\tau)$ of (a,b) Stokes--anti-Stokes and (c) leapfrog emission from OM systems, calculated for the cavity fields $a$ (solid lines), and fluctuations $\delta a$ (dashed lines), with insets to the right of the figure representing spectra $S_\Gamma$ (solid lines) and $S_\Gamma[\delta a]$ (dashed lines). Vertical gray strips in the insets represent the spectral positions and widths $\Gamma$ of the filters, superimposed on the emission spectra. For the Stokes--anti-Stokes correlations, we choose the coupling parameters and filter widths (a) $(g_0, \Gamma)/\kappa=(0.08, 0.05)$ and (b) $(g_0, \Gamma)/\kappa=(1, 0.5)$, and for the leapfrog case we set (c) $(g_0,~\Gamma)/\kappa=(0.2,~0.05)$. We assume the vanishing thermal phonon populations $\nbth=0$, and set all the remaining parameters, including phonon decay rate $\gamma/\kappa=0.1$, to the values listed in the caption of Fig.~\ref{fig:5coupling}.
}
	\label{fig:5temporal}
\end{figure}

In this section, we will consider the dynamics of the frequency-resolved correlations in the Stokes--anti-Stokes and leapfrog emission processes discussed in the previous section. Furthermore, we will also study in more detail the mechanism of interference antibunching previously suggested as an explanation for the Stokes--anti-Stokes correlation behaviour, and briefly comment on the connection between the temporal and spectral resolution of our setup.

\subsubsection{Stokes--anti-Stokes correlation dynamics}
\label{subsec:SaStime}

To illustrate the origin of the interference antibunching of the Stokes--anti-Stokes correlations found for $g_0/\kappa< 0.1$, we plot in Fig.~\ref{fig:5temporal}(a) the time-delayed correlations $g_\Gamma^{(2)}[a](\omega_S,\omega_{aS};\tau)$ (solid orange line) for the parameters that maximize the antibunching found in Fig.~\ref{fig:5coupling}(a): $(g_0, \Gamma)/\kappa=(0.08, 0.05)$. We observe in these time-delayed correlations that the antibunching $g_\Gamma^{(2)}[a](\omega_S,\omega_{aS};\tau=0)<1$ results from the large-amplitude oscillations with frequency $\omega_b$, which are not present when considering correlations of the fluctuations $g_\Gamma^{(2)}[\delta a](\omega_S,\omega_{aS};\tau)$ (dashed red line). This corroborates our previous statement that the interference antibunching results from the elastic component dominating over the anti-Stokes emission line, rather than originating from the Stokes--anti-Stokes emission mediated by a real one-phonon state $\ket{0_a, 1_b}$.

Notably, the strong bunching of the fluctuations in Fig.~\ref{fig:5temporal}(a) is asymmetric near $\tau=0$, and decays as $\exp(-\Gamma |\tau|)$ (with filter response time) for larger delays. Since this emission pathway is mediated by an excited phonon state, we would expect the correlations to decay over time with the characteristic rate of phonon decay ($\gamma$) instead. This discrepancy can be attributed to the choice of very narrow filters $\Gamma/\gamma=0.5, \Gamma/\kappa=0.05$, made to prioritize the spectral resolution of the TPS. This narrow filter linewidth imposes a poor temporal response of the setup, which effectively masks the intrinsic dynamics of the Stokes--anti-Stokes emission pathway. This issue can be solved by choosing a range of parameters such that $\Gamma\sim \omega_b, \kappa\gg \gamma$, which would reduce the spectral selectivity of the detection but could offer insights into the dynamics of such processes. We consider such an arrangement in Fig.~\ref{fig:5temporal}(b), setting wide filters $\Gamma/\kappa=0.5$, and simultaneously increasing the coupling  $g_0/\kappa=1$ to amplify the anti-Stokes emission. We plot the correlations with ($g^{(2)}_\Gamma[a]$, solid orange line), and without the elastic component ($g^{(2)}_\Gamma[\delta a]$, dashed red line).  As previously observed \cite{anderson2018}, $g_\Gamma^{(2)}[a](\omega_S,\omega_{aS};\tau)$ is now strongly asymmetric with respect to delay $\tau$. This reflects the fact that in the regime of near-zero steady-state phonon populations, a phonon-annihilating anti-Stokes emission has to be preceded by a Stokes emission in which that phonon is generated. In other words, the two-photon process occurs with a given temporal order.
Furthermore, the correlations decay approximately as $\exp(-\gamma \tau)$ for $\tau>0$, which demonstrates that the increased time resolution, offered by a broad $\Gamma\gg\gamma$ filter, allows us access to the vibrational dynamics of the system. In Appendix~\ref{ap:filter} we provide additional calculations that illustrate the impact of the filtering linewidth on the frequency-resolved correlations of these systems.

\subsubsection{Leapfrog dynamics}

We can perform a similar analysis for the frequency-resolved correlations between the strongly bunched photons emitted in a leapfrog process. The temporal dynamics of this process (Fig.~\ref{fig:5temporal}(c)) again shows large oscillations resulting from the interference with the elastic peak (solid green line), that are removed if we instead consider the correlations of the fluctuations $\delta a$ (dashed line). The resulting correlations are clearly symmetric with respect to the delay time $\tau$, highlighting the fact that there is no particular time order of the emission, and decay exponentially over time. In principle, since this two-photon transition is not mediated by the emission or absorption of phonons, the decay of correlations over time should be governed by the characteristics of the cavity -- $\kappa$ -- and the temporal resolution of the detection setup determined by $\Gamma$. For the regime of parameter chosen, i.e., $\Gamma/\kappa=0.05$, it is the temporal response of the filter which provides the decay time of the correlations. {As with the Stokes--anti-Stokes correlations, to resolve the intrinsic dynamics of the leapfrog correlations, governed by $\kappa$, one would require much broader filters $\Gamma> \kappa$, and a system where the leapfrog processes would be separated far enough from Stokes and elastic emission lines.}

\section{Cauchy-Schwarz inequality: non-classicality}
\label{subsec:CSI}

\begin{figure}[bt]
	\centering
	\includegraphics[width=\linewidth]{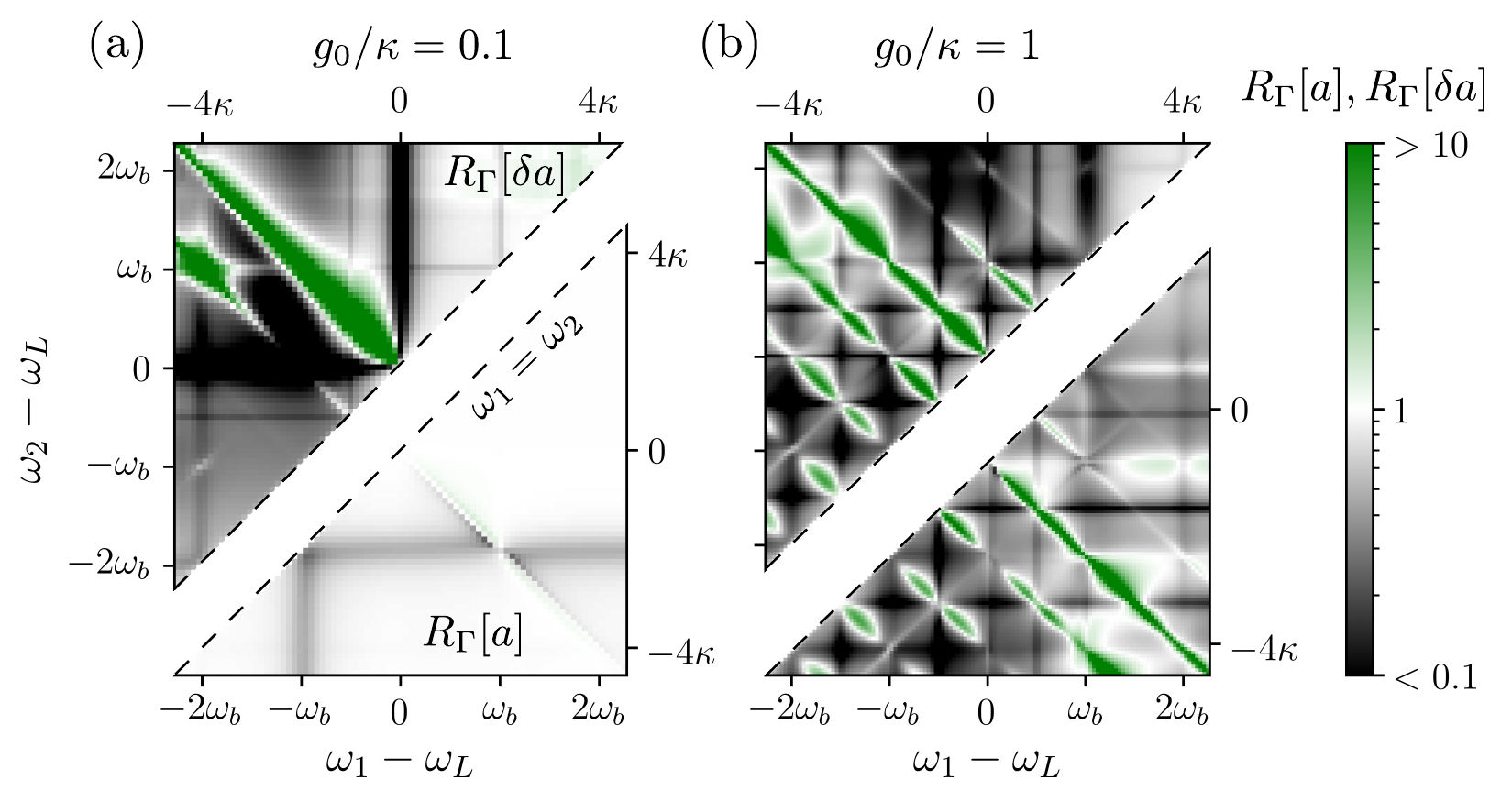}
	\caption{Degree of Cauchy-Schwarz Inequality violation $R_\Gamma[a](\omega_1,\omega_2)$ and $R_\Gamma[\delta a](\omega_1,\omega_2)$ (Eq.~\eqref{CSI}) in an optomechanical system in the (a) weak $g_0/\kappa=0.1$ and (b) strong $g_0/\kappa=1$ single-photon coupling regime. Besides the parameters given in the plots, for all the systems we set the linewidth of the filter as $\Gamma=0.05\kappa$, the energy and decay rate of phonons as $\omega_b/\kappa=2$ and $\gamma/\kappa=0.1$, and consider laser with amplitude $\Omega/\kappa=0.1$ tuned to the first excited eigenstate $\omega_L=\omega_a-\Delta_g$.}
	\label{fig:6}
\end{figure}

Despite the gain of information provided by the frequency-resolved correlations, when compared to their color-blind counterpart these measurements also have an important drawback regarding the ability to unambiguously discern between quantum and classical light fields. While $g^{(2)}(0)<1$ or $g^{(2)}(0)<g^{(2)}(\tau)$ are unambiguous signatures of non-classicality, the observation of $g_\Gamma^{(2)}[a](\omega_1,\omega_2)<1$ or $g_\Gamma^{(2)}[a](\omega_1,\omega_2;0)<g^{(2)}[a](\omega_1,\omega_2;\tau)$ is not. For example, a classical field with phase fluctuations can induce such frequency-resolved antibunching~\cite{silva16a} without the presence of quantum nonlinearities. 

It is, however, possible to test the non-classical character of the correlations between different frequency channels $\omega_1$ and $\omega_2$ by using the information encoded in the TPS $g_\Gamma^{(2)}[a](\omega_1,\omega_2)$ (or in the correlations of any other mode, for example $\delta a$). The idea consists in harnessing one of the features of classical correlations between two random variables, that is, the Cauchy-Schwarz inequality (CSI)~\cite{clauser69a}. Identifying the intensities at the two frequencies with such variables, we can write down the CSI as~
\begin{equation}
\left[g_{\Gamma}^{(2)}[a](\omega_1,\omega_2)\right]^{2} {\leq}~g_{\Gamma}^{(2)}[a](\omega_1,\omega_1)~g_{\Gamma}^{(2)}[a](\omega_2,\omega_2)\,.
\end{equation}

Since the CSI should hold for any classically correlated variables, one can define a parameter (introduced in Ref.~\cite{sanchezmunoz14b}) that indicates the degree of CSI violation:
\begin{equation}\label{CSI}
R_\Gamma[a](\omega_1,\omega_2) = \frac{\left[g_{\Gamma}^{(2)}[a](\omega_1,\omega_2)\right]^{2}}{g_{\Gamma}^{(2)}[a](\omega_1,\omega_1)~g_{\Gamma}^{(2)}[a](\omega_2,\omega_2)},
\end{equation}
yielding a sufficient condition for the observation of non-classical correlations that is $R_\Gamma[a](\omega_1,\omega_2)>1$. Furthermore, we can also measure the degree of non-classicality of the inelastically scattered light, by calculating $R_\Gamma$ from the fluctuations $\delta a$, denoted as $R_\Gamma[\delta a](\omega_1,\omega_2)$. In Figs.~\ref{fig:6}(a,b) we plot the frequency maps of $R_\Gamma[a]$ and  $R_\Gamma[\delta a]$ for the same parameters as in Figs.~\ref{fig:4}(a,c), showing how the strongly bunched regions of the TPS often display a large CSI violation (in green), being therefore a source of non-classical correlations. The results in (a) correspond to $g_0/\kappa=0.1$ and  reveal a strong violation of CSI over a broad region near the $\omega_1+\omega_2=2\omega_L$ antidiagonal. For $g_0/\kappa=1$ (panel (b)) the CSI is violated for narrower regions around multiple $\omega_1+\omega_2=2\omega_L+n\omega_b$ antidiagonals. {As in Fig.~\ref{fig:4}(c), these additional features originate form the intrinsic nonlinearity of the optomechanical Hamiltonian, and are therefore absent in the CSI maps of its linearized form (not shown here).}
{We also find that the violation of CSI can be enhanced if we remove the coherent elastic field which dominates the emission in (a)}. One of the differences with the already observed CSI violation in OM systems~\cite{anderson18a} is that here one does not rely on a heralding preparation step.

\section{Conclusions \& outlook}
\label{sec:conclu}

In summary, we present a systematic study of frequency-resolved correlations in cavity optomechanical systems in the sideband-resolved regime and for moderate to strong single-photon coupling strengths. We show how the two-photon correlation spectra unveil a rich landscape of correlations hidden in other observables. We also provide an intuitive picture that explains these correlations based on the anharmonic level structure and multi-mode squeezing Hamiltonians appearing through the nonlinear optomechanical coupling, and test their non-classical nature based on the Cauchy-Schwarz inequality violation. Importantly, non-trivial frequency-resolved correlations appear already for smaller coupling strengths than for the frequency blind correlations, thus opening new avenues to observe nonlinear phenomena in optomechanical  systems operating far from the single-photon strong coupling limit.

Furthermore, we believe this work opens many research directions that one can follow. For example, although we focused on a particular range of parameters involving low optical quality factors, $Q$'s, and high-frequency optical phonons, which best describes the novel implementation of optomechanics in molecular systems \cite{roelli2014molecular,schmidt16a}, there are many other relevant questions to be answered, e.g., what will be the role of phonon population in systems with low-frequency phonon modes where this population will be non-negligible?; how will the correlation maps change with incoherent pumping?; how will the balance of effects of Kerr and multi-phonon nonlinearities change in larger-Q, weakly pumped cavities? Another interesting direction could be to harness the knowledge acquired through these frequency-resolved correlations to connect it with recent findings in photonic Cooper pairs~\cite{saraiva17a}, to design non-classical photon~\cite{sanchezmunoz14a,sanchezmunoz18a} or, as recently proposed in~\cite{bin19a}, phonon sources.

\section*{Funding Information}

MKS thanks Michael J. Steel for stimulating discussions, and acknowledges funding from Australian Research Council (ARC) (Discovery Project DP160101691) and the Macquarie University Research Fellowship Scheme.  AGT acknowledges support from CSIC Research Platform on Quantum Technologies PTI-001 and from Spanish project PGC2018-094792-B-100 (MCIU/AEI/FEDER, EU). RE and JA acknowledge project PID2019-107432GB-I00 from the Spanish Ministry of Science and Innovation, project H2020-
FET Open “THOR” Nr. 829067 from the European
Commission, and grant IT1164-19 from the Basque Government for consolidated groups of the Basque University.

\appendix

\section{Numerical calculation of spectra and two-photon spectra}\label{sec:methods}
The two-photon spectra were calculated using the method developed originally by del Valle \textit{et al.}~\cite{delvalle12a}. In that work, authors proposed to calculate the spectra and frequency-resolved correlations by coupling the mode of interest --- in this case the cavity mode $a$ --- to additional \textit{sensor} modes with bosonic anihilation operators $\varsigma_i$, characterized by resonant frequencies $\omega_i$, and spontaneous decay rates $\Gamma$:
\begin{equation}\label{eq.sensorcoupling}
    H_{\varsigma} = \sum_{i=1,2} \omega_i \varsigma_i^\dag \varsigma_i + \varepsilon_{i}(\varsigma_i a^\dag+\varsigma_i^\dag a).
\end{equation}
The spectra $S_\Gamma(\omega)[a]$ and frequency-resolved correlations $g_{\Gamma}^{(2)}[a]$ can be then retrieved from the respective steady-state expectation values as
\begin{equation}\label{eq:SGamma}
    S_{\Gamma}(\omega_1)[a] = \lim_{\varepsilon_1\rightarrow 0} \frac{\Gamma}{2\pi \varepsilon_1^2} \mean{\varsigma_1^\dag\varsigma_1},
\end{equation}
\begin{equation}\label{eq:g2Gamma}
    g_{\Gamma}^{(2)}[a](\omega_1,\omega_2) = \lim_{\varepsilon_1,\varepsilon_2\rightarrow 0} \frac{\mean{\varsigma_1^\dag\varsigma_2^\dag\varsigma_1\varsigma_2}}{\mean{\varsigma_1^\dag\varsigma_1}\mean{\varsigma_2^\dag\varsigma_2}}.
\end{equation}
This method can be naturally extended to measure the spectra and correlations of other modes, by simply substituting the mode operator $a$ in Eq.~\eqref{eq.sensorcoupling}. For example, spectrum $S_\Gamma[\delta a]$ and TPS $g_\Gamma^{(2)}[\delta a]$ of cavity mode fluctuations $\delta a$ are found by calculating Eqs.~\eqref{eq:SGamma} and \eqref{eq:g2Gamma} for sensors governed by Hamiltonian
\begin{equation}\label{eq.sensorcoupling2}
    H_{\varsigma,\delta a} = \sum_{i=1,2} \omega_i \varsigma_i^\dag \varsigma_i + \varepsilon_{i}\left[\varsigma_i (a^\dag-\mean{a}^*)+\varsigma_i^\dag (a-\mean{a})\right].
\end{equation}
This calculation is simplified by the observation that, in the limit of vanishing coupling $\varepsilon_i\rightarrow 0$, the detection setup does not perturb the system, and the value of $\mean{a}$ can be calculated in the absence of $H_{\varsigma,\delta a}$.

The procedure proposed in this original work was then simplified following distinct, but ultimately equivalent formulations by L{\'o}pez Carre{\~n}o \textit{et al.} \cite{lopezcarreno18a}, and by Holdaway \textit{et al.}~\cite{holdawaya18a}. In the latter, authors simultaneously solved two inherent difficulties of the original method: ensuring that the backaction from sensors onto the quantum system is vanishingly small while retaining the numerical accuracy of the method, and avoiding increasing the Hilbert space of the system by including the sensors in the quantum system. Our implementation of this algorithm uses the QuTiP toolbox \cite{johansson12a,johansson13a}, and linear algebraic solvers for sparse matrices (scipy.sparse.linalg.spsolve) to calculate the vectorized \textit{auxiliary conditional states} $|\rho_i^j\rrangle $. To calculate the maps of the Cauchy-Schwarz inequality violation, we modified this algorithm to calculate intensity auto-correlations $g_{\Gamma}^{(2)}(\omega_i,\omega_i) = \lim_{\varepsilon_i\rightarrow 0} \mean{(\varsigma_i^\dag)^2\varsigma_i^2}/\mean{\varsigma_i^\dag\varsigma_i}^2$. The expression for $\mean{(\varsigma_i^\dag)^2\varsigma_i^2}$, quadratic in the cavity-sensor coupling parameter $\epsilon_i$, is found by following the recipe detailed in Section IIIA of Ref.~\cite{holdawaya18a} up to Eq.~(17), and --- adopting the notation from that contribution --- defining the following {auxiliary conditional states}: $\rho_0^2$ (in vectorized form denoted as $|\rho_0^2 \rrangle$), $\rho_1^2$ ($|\rho_1^2\rrangle$) and $\rho_2^2$ ($|\rho_2^2\rrangle$) as
\begin{equation}\label{eq:eq02}
|\rho_0^2\rrangle \approx \frac{-i\sqrt{2}\epsilon}{\mathcal{L}_0-(\Gamma-2i\omega_1)}|\rho_0^1\rrangle a^\dag,
\end{equation}
\begin{equation}\label{eq:eq12}
|\rho_1^2\rrangle \approx \frac{i\epsilon}{\mathcal{L}_0-\left(\frac{3}{2}\Gamma-i\omega_1\right)}\left(a|\rho_0^2\rrangle - \sqrt{2}|\rho_1^1\rrangle a^\dag\right),
\end{equation}
\begin{equation}\label{eq:eq22}
|\rho_2^2\rrangle = \frac{i\sqrt{2}\epsilon}{\mathcal{L}_0-2\Gamma}\left(a|\rho_1^2\rrangle - |\rho_2^1\rrangle a^\dag\right),
\end{equation}
and identifying
\begin{equation}
\mean{(\varsigma_i^\dag)^2\varsigma_i^2} = 2\left(\frac{\epsilon_i}{2\pi}\right)^2\Tr\left(\rho_2^2\right).
\end{equation}
For the details of this method and clarification of the notation used, we direct the reader to Ref.~\cite{holdawaya18a}.

We should note that this autocorrelation could equivalently be calculated by considering the cross-correlation between two identical sensors ($\omega_1=\omega_2$ and $\varepsilon_1=\varepsilon_2$) using the algorithm proposed in Ref.~\cite{holdawaya18a}. However, we found that, for each point on the map of CSI, this method requires 9 calls to the scipy.sparse.linalg.spsolve procedure to calculate the vectorized density matrices (one for each of Eqs.~(25a-h) in Ref.~\cite{holdawaya18a}), compared to 5 required for the implementation described above (two for $|\rho_1^0\rrangle$ and $|\rho_1^1\rrangle$ and 3 for Eqs.~(\ref{eq:eq02}-\ref{eq:eq22})).

The Python implementation of this method is available upon request from the corresponding authors.

\section{TPS of two-mode squeezed Hamiltonians}\label{sec:coupled_cavities}

\begin{figure}[!h]
	\centering
	\includegraphics[width=\linewidth]{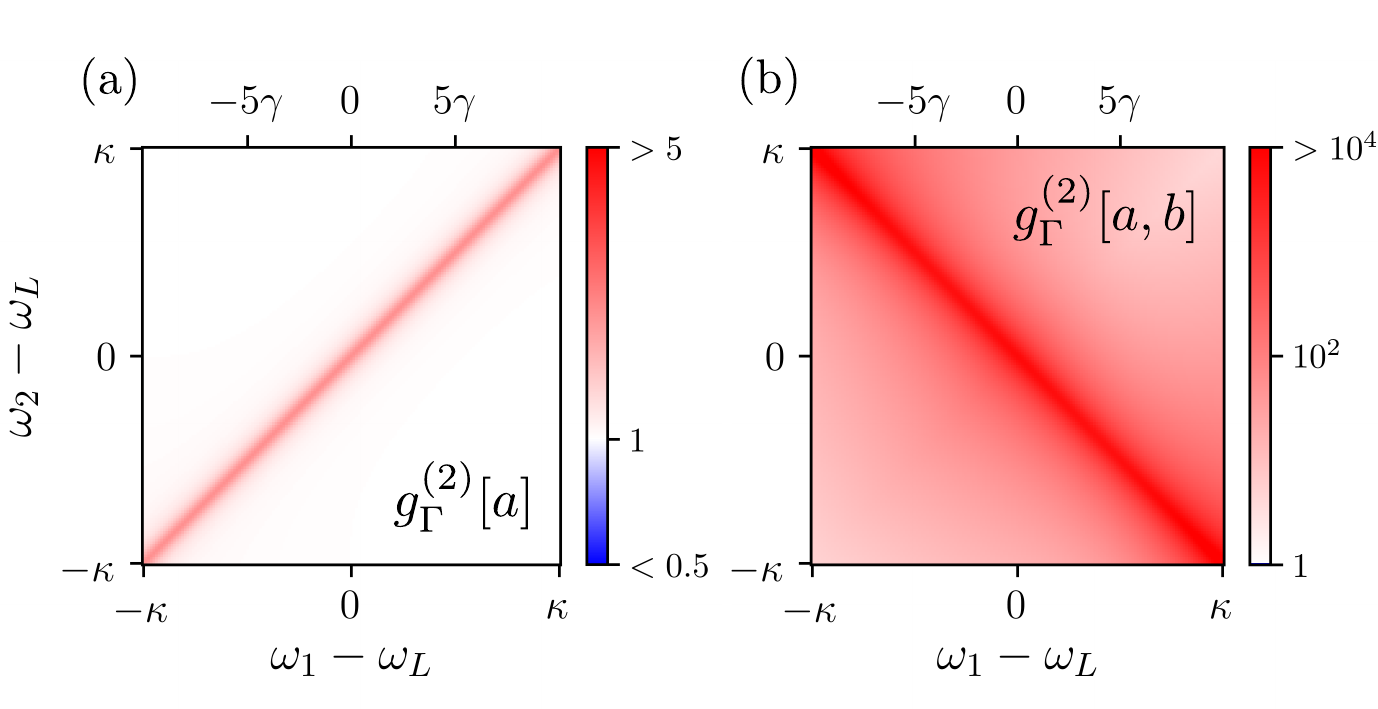}
	\caption{Two-photon spectra of a two-mode squeezing Hamiltonian given in Eq.~\eqref{eq:ham.appendix}, calculated as correlations of (a) mode $a$, (b) modes $a$ and $b$. All parameters match those used in Fig.~\ref{fig:3} of the main text.}
	\label{fig:appendix_b}
\end{figure}

In Section \ref{subsec:antidial}, when analyzing the correlations induced by the term $H^{(1)}$ of the Hamiltonian $\tilde{H}_L$ describing multi-phonon processes (Eq.~\eqref{eq:coh.pumping.transformed.2}), we briefly mentioned that the two-mode squeezing (\textit{active}) terms $ab+a^\dag b^\dag$ drive the systems into strongly correlated states $\ket{i_a,i_b}$, but do not --- in the absence of \textit{passive terms} $a b^\dag+a^\dag b$ --- induce strong autocorrelations of mode $a$. 

To illustrate this mechanism, we consider the Hamiltonian with only the \textit{active terms}
\begin{equation}\label{eq:ham.appendix}
    H=\Delta_a a^\dag a+\omega_b b^\dag b+i\Omega \frac{g_0}{\omega_b}(a^\dag b^\dag-ab),
\end{equation}
and, in Fig.~\ref{fig:appendix_b}(a), plot the TPS of mode $a$: $g_\Gamma^{(2)}[a]$, finding only the diagonal indistinguishability feature. To identify the antidiagonal leapfrog bunching, we show in (b) the two-mode frequency resolved correlation of modes $a$ and $b$ --- $g_\Gamma^{(2)}[a,b]$, calculated from Eq.~\eqref{eq:g2Gamma} using the correlations between the two sensors coupled to the optomechanical system via the Hamiltonian
\begin{equation}\label{eq.sensorcoupling3}
    H_{\varsigma,ab} = \sum_{i=1,2} \omega_i \varsigma_i^\dag \varsigma_i + \varepsilon_{1}\left[\varsigma_1 a^\dag +\varsigma_1^\dag a\right] + \varepsilon_{2}\left[\varsigma_2 b^\dag +\varsigma_2^\dag b\right].
\end{equation}

\section{Dependence of correlations on the width of the filter\label{ap:filter}}
\begin{figure}[!htbp]
	\centering
	\includegraphics[width=\linewidth]{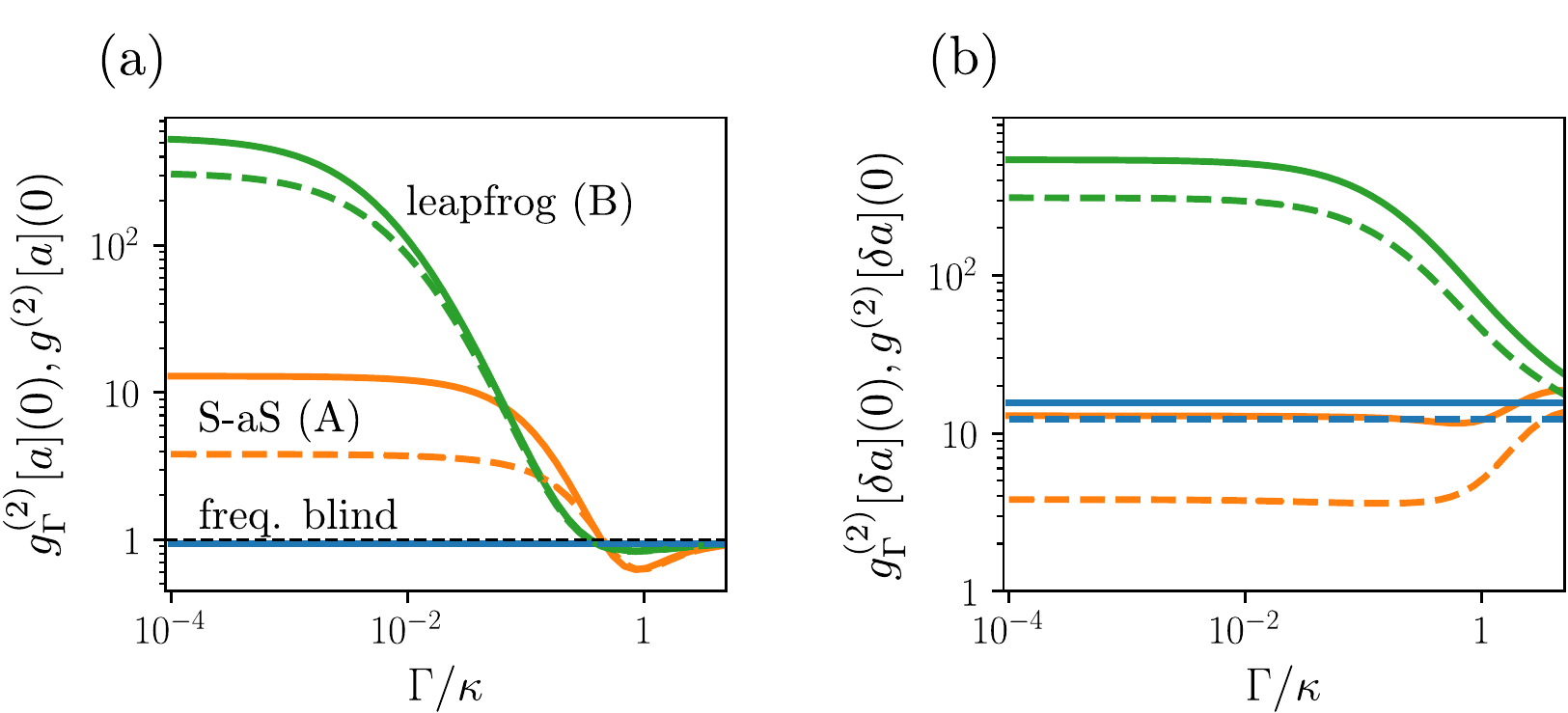}
	\caption{Dependence of the two-photon correlations  on filter width $\Gamma$, calculated for the cavity modes with (a, $g_\Gamma^{(2)}[a](\omega_1,\omega_2;\tau=0)$)  and without (b, $g_\Gamma^{(2)}[\delta a](\omega_1,\omega_2;\tau=0)$) the contribution from the elastic component. In all the cases we consider correlations of
	Stokes--anti-Stokes pairs (orange solid and dashed lines; $\omega_1=\omega_L-\omega_b$, $\omega_2=\omega_L+\omega_b$), leapfrog transitions (green; $\omega_1=\omega_L-0.5\omega_b$, $\omega_2=\omega_L+0.5\omega_b$) and frequency-blind correlations $g^{(2)}(\tau)$ (blue) with the solid and dashed lines denoting calculations assuming vanishing ($\nbth=0$) and non-vanishing ($\nbth=0.1$) thermal populations of phonons, respectively. Besides the parameters given in the plots, for all the systems we set the energy and decay rate of phonons as $\omega_b/\kappa=2$ and $\gamma/\kappa=0.1$, and consider a laser with amplitude $\Omega/\kappa=0.1$ tuned to the first excited eigenstate $\omega_L=\omega_a-\Delta_g$, with coupling $g_0/\kappa=0.5$.}
	\label{fig:appendix_c}
\end{figure}

In Fig.~\ref{fig:appendix_c} we explore the dependence of selected correlations on the width of the filters. In the limit of large $\Gamma$, irrespective of the frequency of the filters, we recover the values of frequency-blind correlations $g^{(2)}$ (shown with blue lines). 

In the limit of narrow filters, the frequency-resolved (Stokes--anti-Stokes and leapfrog) correlations of mode $a$ and its fluctuations become identical as the elastic contribution is filtered out. With the increasing spectral resolution, all information on the time of emission is lost. Interestingly, unlike for the incoherently pumped systems discussed by Gonzalez-Tudela \textit{et al.} \cite{gonzaleztudela13a}, we do not recover the simple general limits of $g_\Gamma^{(2)}(\omega_1,\omega_2)=2$ for $\omega_1=\omega_2$, and $=1$ for $\omega_1\neq\omega_2$. 

\bibliography{bibliography}

\end{document}